# Characterization of the Compact Multiplanet System TOI-789: Validation of Three Terrestrial Planets, Dynamical Characterization, and Detection of a Fourth Temperate Candidate

Juan F. Pagés Navarrete
Independent Researcher, Málaga, Spain; jfpages@gmail.com
Received 2026 June 19; revised 2026 August 10; accepted 2026 September 10; published 2026 October 7

## Abstract

We present the validation and dynamical characterization of TOI-789 (TIC 300710077), a compact multiplanet system around an M3 dwarf at 43 pc. Using ∼7 yr of TESS photometry (42 sectors), public ground-based follow-up (LCOGT, MEarth), and high-resolution imaging (Gemini/Zorro, SOAR), we statistically validate three terrestrial planets with TRICERATOPS: TOI-789 b ($R_p = 1.18\,R_\oplus$, $P = 5.45$ days), c ($1.15\,R_\oplus$, 8.04 days), and d ($1.40\,R_\oplus$, 12.97 days), with false-positive and nearby-false-positive probabilities below the validation thresholds once resolved neighbours are excluded using the on-target ground-based photometry. Through an iterative Transit Least Squares search we identify a fourth, previously uncatalogued planet candidate ($P = 19.74$ days, $R_p \approx 0.88\,R_\oplus$), the outermost and most temperate member ($S = 1.71\,S_\oplus$, $T_{eq} \approx 292$ K); its low signal-to-noise ratio precludes formal validation, so we report it as a candidate pending confirmation. A dynamical analysis (SPOCK and $N$-body integrations) indicates a dynamically cold, near-resonant system that is stable only for low eccentricities ($e \lesssim 0.05$); it is eccentricity rather than planetary mass that limits the system's stability, and including the fourth candidate tightens this constraint further. The four objects span the Earth–Venus transition across the Venus zone, with two validated planets falling in the temperate regime under the equilibrium-temperature criterion, making the system a potential comparative-planetology laboratory anchored on confirmed planets. The low TSM (∼3–5) and moderate host brightness place atmospheric characterization beyond reach. TOI-789 exemplifies the numerically dominant population of compact terrestrial-planet systems around M dwarfs whose characterization is limited by host brightness for which dynamical characterization offers a complementary route.

*Unified Astronomy Thesaurus concepts:* Exoplanet systems (484); Transit photometry (1709); Exoplanet dynamics (490); M dwarf stars (982); Habitable zone (696)

## 1. Introduction

A central question in exoplanet science is what determines the habitability of a terrestrial planet. In the solar system, Earth and Venus—bodies of similar mass, size, and composition—followed radically different evolutionary paths: while Earth maintained temperate conditions, Venus evolved into a runaway-greenhouse state, with a dense $CO_2$ atmosphere and extreme surface temperatures (S. R. Kane et al. 2014). Identifying the insolation threshold that separates these two fates—the inner edge of habitability—is essential for interpreting the diversity of temperate worlds now being discovered (R. K. Kopparapu et al. 2014). In this context, multiplanet systems hosting several terrestrial planets around a single star, subject to increasing insolation, provide natural comparative-planetology laboratories in which the stellar properties are fixed and only the received irradiation varies.

M dwarfs are the ideal hosts for this search. Their small radii amplify the transit depths of small planets, and their low luminosities shift the temperate and habitable zones to short orbital periods accessible to transit surveys (C. D. Dressing & D. Charbonneau 2015; A. L. Shields et al. 2016). Missions such as Kepler and TESS (G. R. Ricker et al. 2015) have revealed that small planets are abundant around M dwarfs (K. Ment & D. Charbonneau 2023; E. D. Gillis et al. 2026) and that they frequently organize into compact, dynamically cold, multiplanet systems with uniform radii—the so-called "peas in a pod" pattern: within a single system, planets have radii (and masses) far more similar to one another than expected from random draws of the global population, a regular orbital spacing, and a mild tendency for the outer planet to exceed its inner neighbor in size, interpreted as the imprint of self-regulated formation (S. Millholland et al. 2017; L. M. Weiss et al. 2018). Systems such as TRAPPIST-1 (M. Gillon et al. 2017), TOI-270 (M. N. Günther et al. 2019) and L 98-59 (V. B. Kostov et al. 2019) exemplify this architecture and have become benchmark targets for the study of temperate terrestrial planets.

However, only a small fraction of these systems orbit stars bright and nearby enough to permit radial-velocity mass measurements or atmospheric characterization. Of the ∼980 planets that TESS is expected to detect around M dwarfs, only of order 50 orbit hosts bright enough ($J < 9.5$) for detailed follow-up (P. W. Sullivan et al. 2015), and JWST atmospheric characterization is essentially restricted to the nearest M dwarfs (∼10–15 pc; F. Wunderlich et al. 2019). Most systems therefore reside around fainter hosts for which such measurements are inaccessible with current instrumentation. For these systems, statistical validation and dynamical characterization emerge as the principal routes to constraining their properties. Statistical validation in particular has now been applied to the TESS Full Frame Images at survey scale, newly validating over a hundred planets and releasing thousands of vetted candidates (e.g., M. Lafarga et al. 2026). In particular, orbital-stability analysis yields information independent of mass measurements: requiring that a multiplanet system have survived over the lifetime of its star imposes constraints on the masses and, especially, on the eccentricities of its components (J. Fang & J.-L. Margot 2013; B. Pu & Y. Wu 2015; D. Tamayo et al. 2017). This approach is especially valuable for systems that escape spectroscopic characterization. The development of machine learning stability classifiers such as SPOCK (D. Tamayo et al. 2020) has made it computationally feasible to explore the dynamical parameter space of these systems exhaustively.

In this work, we present the validation and characterization of TOI-789 (TIC 300710077), an M3 dwarf at 43 pc. The system was alerted by TESS with three-planet candidates (TOI-789.01, .02, and .03), which we validate statistically here. In addition, taking advantage of the mission's long temporal baseline (∼7 yr, 42 sectors), we identify a fourth, previously uncatalogued transit signal at $P \approx 19.74$ days, which constitutes a new planet candidate—the outermost and most temperate member of the system—detectable thanks to the larger number of transits accumulated over the extended mission. We give particular attention to characterizing the architecture and dynamical stability of the system, which we use both to constrain eccentricities and masses and to assess the dynamical viability of this fourth candidate. Finally, we place the system in the context of the temperate zone and habitability. The paper is organized as follows. Section 2 describes the observations; Section 3, the stellar parameters; Section 4, the validation and vetting; Section 5, the photometric analysis; Section 6, the architecture and dynamical stability; and Section 7, the habitability and characterization prospects. Sections 8 and 9 present the discussion and conclusions.

## 2. Observations

### 2.1. TESS Photometry

TOI-789 (TIC 300710077) was observed by the Transiting Exoplanet Survey Satellite (TESS; G. R. Ricker et al. 2015) at 2 minute cadence across 42 sectors (sectors 1–3, 5–13, 27–33, 35–39, 61–67, 69, 87–90, and 93–98), spanning 2018–2025 and providing a temporal baseline of about 7 yr. We use the light curves processed by the Science Processing Operations Center (SPOC; J. M. Jenkins et al. 2016), adopting the PDCSAP flux (J. C. Smith et al. 2012; M. C. Stumpe et al. 2014) for the transit analysis. For the search for rotational modulation (Section 4.4), we additionally use the SAP flux, which better preserves low-frequency stellar variability. The data were downloaded and processed with the `lightkurve` package (Lightkurve Collaboration 2018), discarding points with poor-quality flags. The SPOC aperture contains, besides the target, three contaminating sources (Figure 1): TIC 300710075 ($\Delta m \approx 2.6$ mag), which dominates the contamination with ∼9% of the flux, and two much fainter sources (TIC 300710083 and TIC 764821353, $\Delta m > 6$) whose combined contribution is negligible (<0.4%). The SPOC pipeline estimates a mean contamination factor CROWDSAP ≈ 0.855 which, incorporating all sources and the diffuse background via the TESS pixel response function (PRF), indicates that 85.5% of the flux in the aperture comes from the target. We correct the transit depths for this dilution, which increases the planetary radii by 8.2% relative to the uncorrected values.

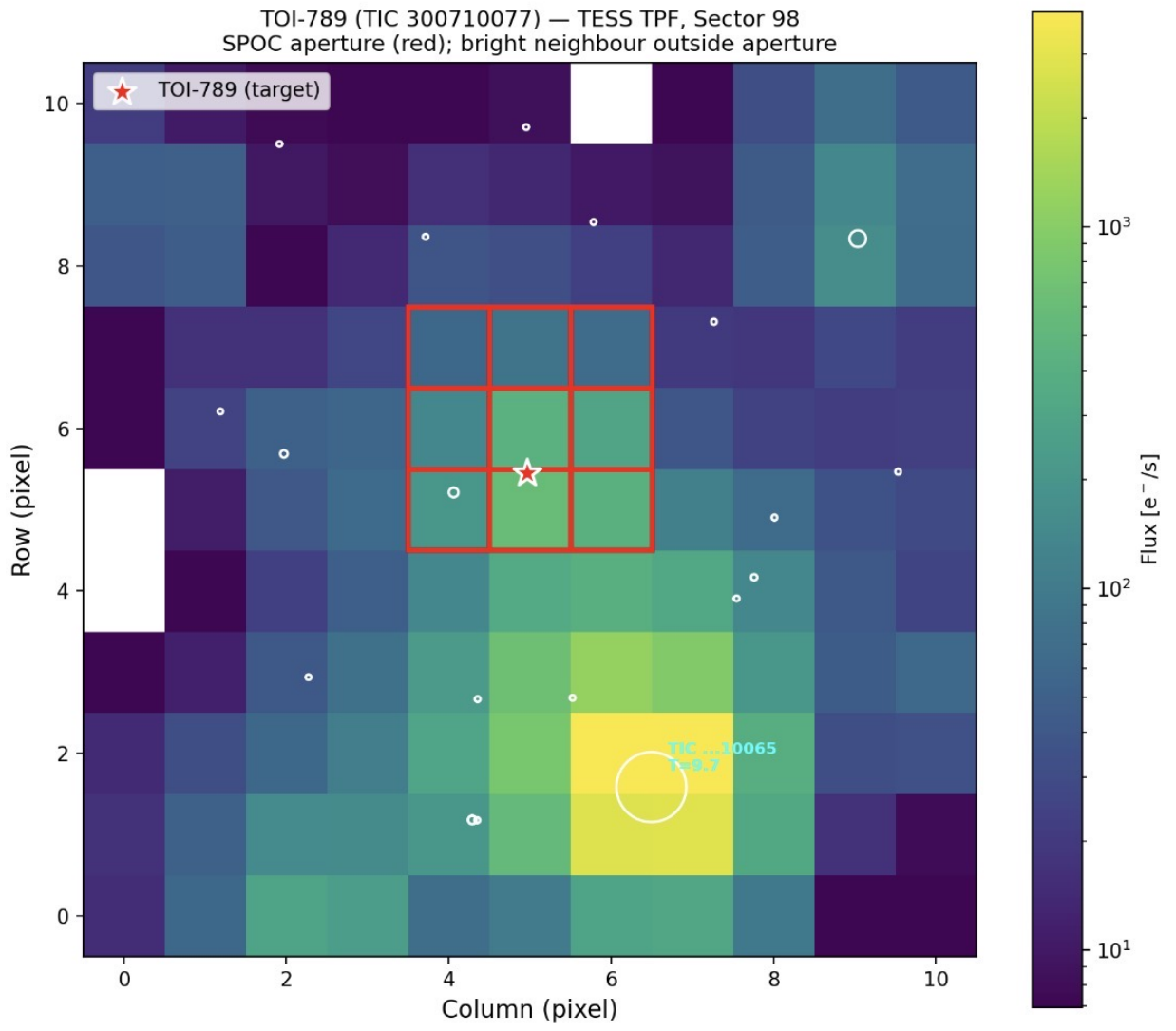

**Figure 1.** TESS target pixel file (TPF) of TOI-789 (TIC 300710077) for Sector 98, with flux per pixel shown on a logarithmic color scale ($e^{-}$ $s^{-1}$). The SPOC photometric aperture is outlined in red and the target is marked with a star; open circles denote Gaia DR3 sources, with symbol size scaled to brightness. The only bright source in the field, the neighbor TIC 300710065 ($T = 9.7$), lies ∼80″ to the southeast, well outside the aperture and therefore unable to be the origin of the transit signals. The contaminant TIC 300710075 ($\Delta m \approx 2.6$) falls within the aperture and is responsible for most of the dilution corrected in Section 2.1 (CROWDSAP ≈ 0.855).

### 2.2. Ground-based Follow-up Photometry

The system has been the subject of extensive ground-based photometric follow-up coordinated by the TESS Follow-up Observing Program (TFOP), whose products are publicly available in the ExoFOP-TESS archive. These observations were obtained with the Las Cumbres Observatory Global Telescope network (LCOGT; T. M. Brown et al. 2013) and MEarth-South (J. M. Irwin et al. 2015). Despite the shallow transit depths (∼1 ppt), multiple campaigns in the red ($i'$) band recovered the events of b ($P = 5.45$ days), c ($P = 8.04$ days), and d ($P = 12.97$ days) on-target in uncontaminated 5″–7″ apertures, with depths and durations consistent with those predicted within the quality of each night, confirming that the signals originate from the target star and not from a neighbor. These observations are further used to rule out background eclipsing binaries among neighboring stars (NEBs) that could reproduce the shallow TESS signals through blending in its large pixels (see M. Kunimoto et al. 2025).

### 2.3. High-resolution Imaging

High-resolution speckle imaging is available from the Zorro instrument on the 8 m Gemini-South telescope (N. J. Scott et al. 2021), in the 562 and 832 nm bands, as well as speckle observations from the 4.1 m SOAR telescope in the $I$ band (A. Tokovinin 2018). Neither reveals a stellar companion (Figure 2). Zorro reaches a contrast of Δmag ≈ 6.8 at 0.″5 in the 832 nm band; SOAR, with its smaller aperture, sets a shallower limit ($\Delta$mag $\approx$ 3.7 at 0$''$.5, reaching $\Delta$mag $\approx$ 6.5 at $\sim$3$''$). These limits exclude unresolved companions that could be the source of the transit signals or dilute their depth.

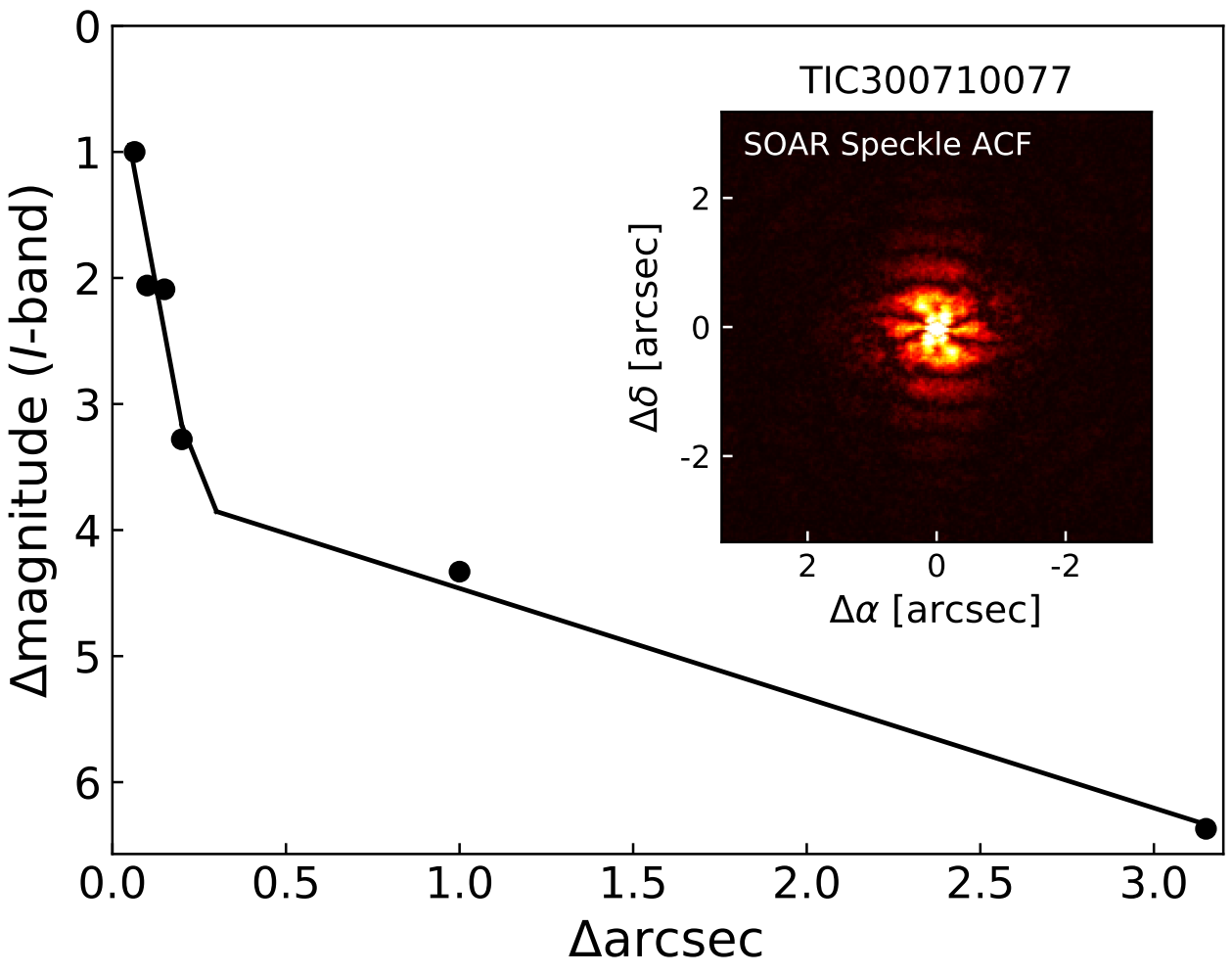

**Figure 2.** SOAR/HRCam $I$-band $5\sigma$ contrast curve for TOI-789, with the speckle autocorrelation function shown in the inset. No stellar companion is detected; the limit ($\Delta$mag $\approx$ 3.7 at 0$''$.5, reaching $\approx$6.5 at $\sim$3$''$) excludes unresolved companions that could be the source of, or dilute, the transit signals. The deeper Gemini-South/Zorro 832 nm limit ($\Delta$mag $\approx$ 6.8 at 0$''$.5; Section 2.3) is the contrast constraint adopted in the TRICERATOPS analysis (Section 4.2).

## 3. Stellar Parameters

TOI-789 (Gaia DR3 5264306681309492864, TIC 300710077) is an M dwarf located at 43.41 $\pm$ 0.06 pc (C. A. L. Bailer-Jones et al. 2021), with a total proper motion of $\sim$196 mas yr$^{-1}$. Its apparent magnitudes are $G = 13.13$, $T = 11.98$, $J = 10.51$, and $K_s = 9.70$.

We derive the fundamental stellar parameters using the empirical, model-independent relations of A. W. Mann et al. (2015, 2019), which for M dwarfs are preferable to model-dependent methods—such as spectral energy distribution (SED) fitting—that suffer from known inconsistencies in this regime (A. W. Mann et al. 2019). From $K_s$ (2MASS) and the Gaia parallax, we obtain an absolute magnitude $M_{K_s} = 6.52 \pm 0.02$, within the calibration range ($4.6 < M_{K_s} < 9.8$). This yields a stellar radius $R_\star = 0.371 \pm 0.011\,R_\odot$ (3.0% precision) and a mass $M_\star = 0.358 \pm 0.009\,M_\odot$, consistent with the TESS Input Catalog (TIC v8.2; K. G. Stassun et al. 2019).

We adopt $T_{\rm eff} = 3471 \pm 64$ K (TIC), corresponding to an approximate spectral type M3 V. Regarding metallicity, no dedicated spectroscopy is available, and the star falls outside the coverage of the large spectroscopic surveys (LAMOST, APOGEE). Gaia DR3 provides two discrepant estimates ([M/H] $= -0.55$ from GSP-phot and $-1.25$ from GSP-spec); both are unreliable since the star ($T_{\rm eff} \approx 3400$ K) lies below the Gaia calibration range ($T_{\rm eff} > 3800$ K; R. Andrae et al. 2023) and the GSP-spec solution yields a $\log g = 3.6$ incompatible with the M-dwarf nature in addition to unfavorable quality flags. We therefore adopt solar metallicity with a conservative uncertainty, [Fe/H] $=0.0 \pm 0.5$ dex. The adopted stellar parameters are summarized in Table 1.

**Table 1**
Stellar Parameters of TOI-789

| Parameter | Value | Source |
|---|---|---|
| TIC ID | 300710077 | K. G. Stassun et al. (2019) |
| Gaia DR3 source ID | 5264306681309492864 | Gaia DR3 |
| R.A. (J2016) | 07:41:04.85 | Gaia DR3 |
| Decl. (J2016) | −71:18:13.55 | Gaia DR3 |
| Distance (pc) | 43.41 $\pm$ 0.06 | C. A. L. Bailer-Jones et al. (2021) |
| $\mu_{\rm tot}$ (mas yr$^{-1}$) | 196.4 $\pm$ 0.1 | Gaia DR3 |
| RUWE | 1.18 | Gaia DR3 |
| $T$ (mag) | 11.98 $\pm$ 0.02 | K. G. Stassun et al. (2019) |
| $G$ (mag) | 13.13 $\pm$ 0.01 | Gaia DR3 |
| $J$ (mag) | 10.51 $\pm$ 0.02 | 2MASS |
| $K_s$ (mag) | 9.70 $\pm$ 0.02 | 2MASS |
| $T_{\rm eff}$ (K) | 3471 $\pm$ 64 | TIC v8.2 |
| Spectral type | $\sim$M3 V | This work |
| $R_\star$ ($R_\odot$) | 0.371 $\pm$ 0.011 | A. W. Mann et al. (2015) |
| $M_\star$ ($M_\odot$) | 0.358 $\pm$ 0.009 | A. W. Mann et al. (2019) |
| $\log g$ (cgs) | 4.85 $\pm$ 0.03 | This work |
| [Fe/H] (dex) | 0.0 $\pm$ 0.5 (adopted) | This work |

**Note.** Radius and mass from the empirical $M_{K_s}$ relations of A. W. Mann et al. (2015, 2019). Metallicity adopted as solar with a conservative uncertainty; the Gaia DR3 GSP-spec/GSP-phot values are unreliable for this cool dwarf ($T_{\rm eff}$ below the calibration range; R. Andrae et al. 2023).

## 4. Transit Search, Validation, and Vetting

### 4.1. Detection and Ephemeris Refinement

The signals of planets b, c, and d were originally identified by the SPOC pipeline (TOI-789.01, .03, .02). We recovered them independently using a Box Least Squares periodogram (BLS; G. Kovács et al. 2002) on the combined light curve of the 42 sectors, after flattening the photometry by masking the transits. We refined the ephemerides taking advantage of the $\sim$7 yr baseline (Section 5).

### 4.2. Statistical Validation

We assess the planetary nature of the signals with the TRICERATOPS package (S. Giacalone & C. D. Dressing 2020; S. Giacalone et al. 2021), which computes the false-positive probability (FPP) and the nearby false-positive probability (NFPP) by comparing the Bayesian likelihoods of different astrophysical scenarios (transiting planet, eclipsing binary, hierarchical systems, and contaminating neighbors), incorporating the simulated background stellar population from TRILEGAL, the SPOC photometric aperture, and the Gemini/Zorro 832 nm contrast curve as a constraint on unresolved companions (S. Giacalone et al. 2021; K. Kawauchi et al. 2022). Following the recommended practice (S. Giacalone et al. 2021; J. Gomez Barrientos et al. 2025; M. Greklek-McKeon et al. 2026), we run the calculation 20 times per planet and report the median and dispersion.

An initial calculation including all cataloged neighbors returns low false-positive probabilities (FPP = 0.005–0.041 for b, c, d) but nonnegligible nearby false-positive probabilities (NFPP$_b$ = 0.002, NFPP$_c$ = 0.037, NFPP$_d$ = 0.011), all exceeding the NFPP $< 10^{-3}$ validation threshold. As an additional check against contamination from neighbors, the SPOC difference-image centroiding shows no significant photocenter offset during the transits, consistent with an on-

**Table 2**
Statistical Validation (TRICERATOPS)

| Planet | FPP | NFPP | FPP × mult. | Disposition |
|---|---|---|---|---|
| TOI-789 b | 0.0030 ± 0.0004 | $<10^{-5}$ | $\sim 6 \times 10^{-5}$ | Validated |
| TOI-789 c | 0.0042 ± 0.0005 | $1.5 \times 10^{-4}$ | $\sim 8 \times 10^{-5}$ | Validated |
| TOI-789 d | 0.0035 ± 0.0010 | $<10^{-5}$ | $\sim 7 \times 10^{-5}$ | Validated |
| TOI-789.04 | 1.0 (robust) | 0.44 | 0.020 | Candidate (low SNR) |

**Note.** Median and dispersion over 20 TRICERATOPS runs (5 for the candidate .04), using a TRILEGAL background population, the SPOC aperture, and the Gemini/Zorro 832 nm contrast curve. Resolved neighbors excluded as transit sources by the on-target ground-based photometry (Section 2.2) are removed from the scenario calculation; with all neighbors included, the NFPP values for b, c, and d are 0.002, 0.037, and 0.011, respectively. The multiplicity boost following J. J. Lissauer et al. (2012) is listed for reference but is not required here. Validation requires FPP < 0.015 and NFPP $< 10^{-3}$ (S. Giacalone et al. 2021), both satisfied by b, c, and d. The candidate .04 is not validated: even after the same neighbor removal it retains FPP = 1.0 and a high NFPP, owing to its low transit SNR ($\sim$2; see M. Kunimoto et al. 2025).

target origin of the signals and supporting the low NFPP values. This NFPP is dominated by the in-aperture contaminant TIC 300710075 ($\Delta m \approx 2.6$ at 19″) and the bright neighbor TIC 300710065 ($T \approx 9.7$ at 80″). Both stars are, however, directly excluded as the source of the transits by the ground-based follow-up: the events of b, c, and d are recovered on-target, at the predicted depth, in uncontaminated apertures of 3.″9–4.″3 radius (Section 2.2), which resolve and exclude these neighbors. Following the TRICERATOPS framework, we therefore remove from the scenario calculation the resolved neighbors ruled out by this seeing-limited photometry and recompute the probabilities. This use of higher-angular-resolution ground-based photometry to resolve and exclude neighbors blended in the TESS aperture, thereby reducing the NFPP, follows the approach demonstrated by J. Gomez Barrientos et al. (2025).

After this vetting, we obtain $\mathrm{FPP}_b = 0.0030 \pm 0.0004$, $\mathrm{FPP}_c = 0.0042 \pm 0.0005$, and $\mathrm{FPP}_d = 0.0035 \pm 0.0010$, with $\mathrm{NFPP} \leqslant 1.5 \times 10^{-4}$ in all three cases (Table 2). All three planets thus satisfy *both* validation criteria of S. Giacalone et al. (2021) simultaneously—FPP < 0.015 and NFPP $< 10^{-3}$, the same thresholds recently adopted by J. Gomez Barrientos et al. (2025) and M. Greklek-McKeon et al. (2026)—without recourse to the multiplicity prior. As an additional, independent reinforcement, the markedly reduced false-positive probability of signals in a multiplanet system (J. J. Lissauer et al. 2012; J. F. Rowe et al. 2014; N. M. Guerrero et al. 2021) lowers the corrected FPPs by a further factor of $\sim$50 (to $\sim 5$–$8 \times 10^{-5}$; Table 2). We conclude that TOI-789 b, c, and d are statistically validated.

### 4.3. A Fourth Planet Candidate

An iterative Transit Least Squares (TLS) search M. Hippke & R. Heller 2019)—masking the detected signals and rerunning the search, following the methodology of R. Heller et al. (2019) for K2-32—revealed a significant periodic signal at $P \approx 19.74$ days (SDE = 24.5), not previously cataloged (Figure 3). We verified its robustness by rerunning the search with a more aggressive masking of planet d (whose period is close to a 3:2 ratio with the signal); the peak persists (SDE = 24.5), ruling out an alias or a residual of the masking.

The signal passes the photometric vetting: a centered transit shape, a consistent odd/even-transit test ($0.6\sigma$), and a negligible fraction of grazing solutions. A juliet fit yields $R_p = 0.81$ (0.6–1.1) $R_\oplus$ before the dilution correction (0.88 $R_\oplus$ after applying the +8.2% correction of Section 5; the value adopted in Table 3), with the uncertainty dominated by the

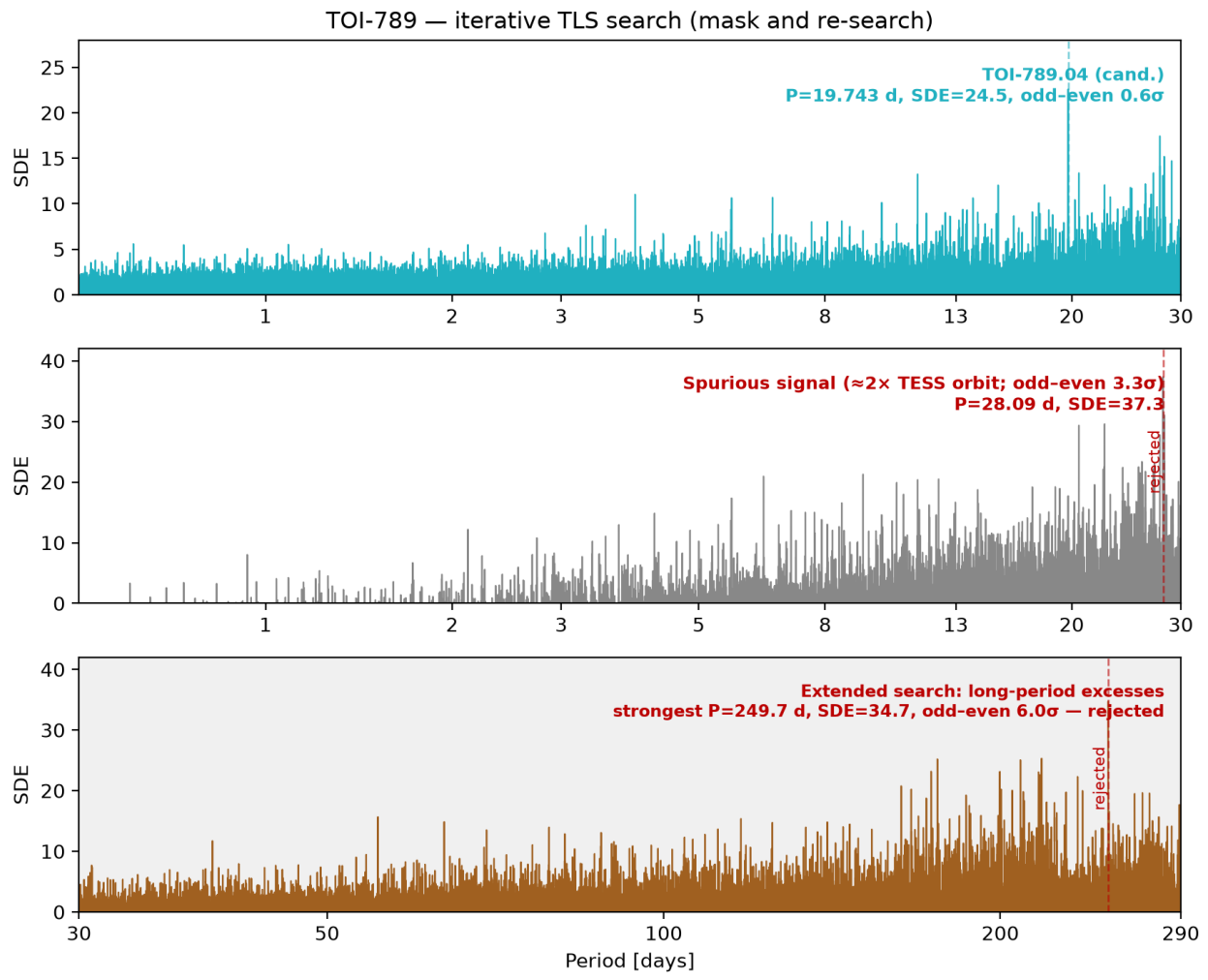

**Figure 3.** Iterative TLS search, masking the signals already identified and rerunning the search (M. Hippke & R. Heller 2019, see R. Heller et al. 2019). The signal detection efficiency (SDE) is plotted against trial period on a logarithmic scale. Top: after masking the three validated planets, a significant peak appears at $P = 19.742$ days (SDE = 24.5), which we report as the candidate TOI-789.04. Middle: after additionally masking the candidate, the strongest remaining peak ($P = 28.09$ days) is spurious—it coincides with $\sim 2\times$ the TESS orbital period and shows a $3.3\sigma$ odd–even mismatch—and is rejected. The top and middle panels share an identical period range (0.5–30 days). Bottom: the search extended to $P = 290$ days (masking all four objects); the only excesses are long-period artifacts, the strongest at $P = 249.7$ days (SDE =34.7, odd–even $6.0\sigma$), all rejected as described in Section 4.4. Dashed lines mark the periods of interest.

radius–impact-parameter degeneracy. Injection–recovery tests confirm that a signal with these parameters is recoverable with TLS (SDE $\approx$ 20), consistent with the detection (Figure 4). We note that the $\sim$19.7 day period of this candidate falls outside the period windows of the uniform short-period transit searches that drive much of the current TESS candidate catalog (e.g., the 0.5–16 days RAVEN search of M. Lafarga et al. 2026), underscoring that the extended multisector baseline of a dedicated, system-specific analysis can recover temperate outer signals that all-sky short-period searches are not designed to detect.

However, the depth ($\sim$400 ppm) corresponds to a low signal-to-noise ratio (SNR $\lesssim$ 2 per point). For reference, M. Kunimoto et al. (2025) chose not to validate the third planet of the HD 101581 system despite an SNR of 7.9—considerably higher than that of our candidate—deeming it insufficient. Applying to the candidate the same neighbor-

**Table 3**
Planetary Parameters of the TOI-789 System

| Parameter | *b* | *c* | *d* | .04 (cand.) |
|---|---|---|---|---|
| TOI designation | .01 | .03 | .02 | .04 (new) |
| Status | Validated | Validated | Validated | Candidate |
| Fitted parameters | | | | |
| $P$ (day) | 5.4470340(50) | 8.0432344(123) | 12.9706890(153) | 19.74323(13) |
| $T_0$ (BJD−2457000) | 1329.1067(13) | 1329.9785(31) | 1334.7672(25) | 1343.2215(147) |
| $R_p/R_\star$ | 0.0265 ± 0.0009 | 0.0263 ± 0.0009 | 0.0310 ± 0.0011 | 0.0201 ± 0.0039 |
| Impact parameter $b$ | 0.27 ± 0.14 | 0.34 ± 0.13 | 0.24 ± 0.14 | 0.71 ± 0.15 |
| Derived parameters | | | | |
| $R_p$ ($R_\oplus$) | 1.18 ± 0.05 | 1.15 ± 0.05 | 1.40 ± 0.06 | 0.88 (0.65–1.2) |
| $a$ (au) | 0.043 ± 0.001 | 0.056 ± 0.002 | 0.077 ± 0.002 | 0.102 ± 0.003 |
| $a/R_\star$ | 25.1 ± 0.7 | 32.6 ± 0.9 | 44.8 ± 1.3 | 58.8 ± 1.7 |
| $S$ ($S_\oplus$) | 9.5 ± 0.9 | 5.7 ± 0.5 | 3.0 ± 0.3 | 1.71 ± 0.16 |
| $T_{eq}$ (K) | 448 ± 11 | 394 ± 9 | 336 ± 8 | 292 ± 7 |
| Mass upper limit ($M_\oplus$, 95%) | <3.7 | <3.0 | <6.6 | ⋯ |
| Eccentricity (dynamical) | <0.05 | <0.05 | <0.05 | <0.05 |
| Shared system parameters | | | | |
| $\rho_\star$ (g cm$^{-3}$) | | 10.1 ± 0.9 | | |
| $q_1$ (Kipping) | | 0.69 ± 0.25 | | |
| $q_2$ (Kipping) | | 0.30 ± 0.23 | | |
| $u_1$, $u_2$ (quadratic) | | 0.50, 0.33 | | |

**Note.** Digits in parentheses give the 1$\sigma$ uncertainty on the final digits. Equilibrium temperatures assume a Bond albedo of 0.3 and full heat redistribution (K. G. Stassun et al. 2019); with zero albedo, values would be ∼10% higher. Planet labels follow orbital-period order (b, c, d), differing from the TOI numbering. Mass upper limits from a TTVFast (K. M. Deck et al. 2014) analysis; eccentricity limits from the dynamical stability analysis (Section 6.2). The tabulated $R_p/R_\star$ is the directly fitted value from the undiluted light curve, whereas $R_p$ ($R_\oplus$) includes the dilution correction ($R_p = 1.082\ (R_p/R_\star)\ R_\star$; + 8.2%, CROWDSAP =0.855; Section 2.1); the two therefore differ by this factor.

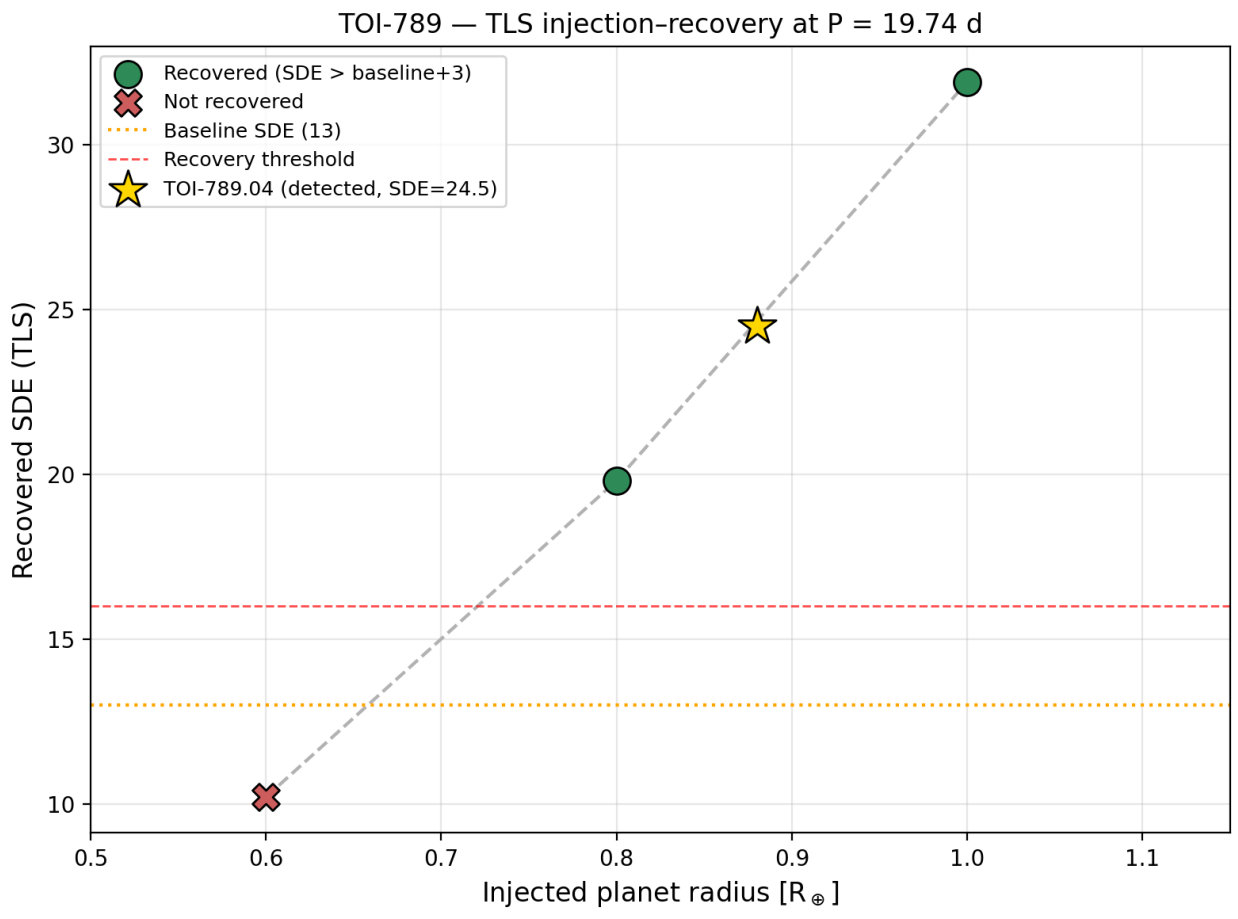

**Figure 4.** TLS injection–recovery test at the candidate period of $P$ = 19.74 days. Synthetic transits of increasing planetary radius were injected into the detrended light curve and recovered with the same pipeline; the recovered SDE is plotted against injected radius. The dotted line marks the baseline SDE and the dashed line the adopted recovery threshold. The candidate TOI-789.04 (star, SDE =24.5) lies well above the threshold, confirming that a transit signal of its inferred size is robustly recoverable and that the nondetection of smaller radii is consistent with the noise level.

removal vetting used for the validated planets (Section 4.2), TRICERATOPS still returns an individual FPP = 1.0 and NFPP ≈ 0.44: unlike for b, c, and d, removing the resolved neighbors does not bring the candidate below the validation thresholds, confirming that its nonvalidation is driven by the low SNR of the signal itself rather than by contamination. Although the multiplicity factor would reduce the FPP, the signal does not reach the SNR threshold (>10) that the multiplicity-validation framework requires to avoid false positives from instrumental systematics (J. F. Rowe et al. 2014; J. L. Christiansen et al. 2020; J. M. Almenara et al. 2024). We therefore do not consider this fourth object validated and report it as a planet candidate (TOI-789.04) pending confirmation. If real, it would be the outermost and most temperate member of the system (insolation ≈1.71 $S_\oplus$; Section 7).

An independent line of support comes from empirical, population-level models of multiplanet architectures. E. V. Turtelboom et al. (2025) applied DYNAMITE (J. Dietrich & D. Apai 2020; J. Dietrich et al. 2022) to the 2024 sample of 183 TESS multiplanet systems—including TOI-789, which entered their sample with its three alerted candidates—to predict the period and radius of an additional, as-yet-undetected planet in each system. For TOI-789, their period-ratio model predicts a fourth planet at $P = 23.2^{+10.2}_{-6.4}$ days with $R_p = 1.11\ R_\oplus$, a period consistent at 1$\sigma$ with our candidate .04 ($P$ = 19.74 days) and a radius compatible within its uncertainties. We stress that this agreement is only suggestive: the prediction is a mechanical output of the model rather than a detection, its period posterior is broad, and the alternative clustered-period model instead favors a short-period planet ($P \approx 3.2$ days) with no counterpart in our data. The period-ratio model predicts the next planet from the regular spacing of adjacent periods in log space (G. D. Mulders et al. 2018), the same intrasystem uniformity that TOI-789 exhibits—candidate .04 continues the sequence near a 3:2 ratio with d ($P_{.04}/P_d = 1.52$; Section 6.1). Thus, its prediction and our detection rest on the same architectural regularity seen directly in the system. Nonetheless, that an empirical model calibrated on the Kepler population independently favors an additional planet near the period of our candidate adds qualitative weight to its planetary interpretation.

### 4.4. Additional Vetting and Rejection of False Positives

For b, c, and d, the differences between the odd- and even-transit depths are $0.99\sigma$, $0.75\sigma$, and $2.00\sigma$, respectively, all below the $3\sigma$ threshold, ruling out eclipsing binaries with twice the period. We detect no significant secondary eclipses. Inspection of the target pixel file confirms that the only bright source ($T_{\rm mag} < 12$) in the field is the target star; the neighbor TIC 300710065 ($T_{\rm mag} \approx 9.7$) lies $\sim 80''$ away, outside the aperture. Archival images from the Digitized Sky Survey (1984–1997), exploiting the star's high proper motion, confirm the absence of background sources at its current position (Figure 5). The neighbor-star analysis (NEBCHECK) on the LCOGT photometry rules out the neighbors as the source of the signals; the few not cleared in individual observations were so because of insufficient precision that night (NEBdepth/rms $< 3$) and not because of the presence of a signal, and they are excluded by combining the on-target detections in uncontaminated apertures (Section 2.2) with the high-resolution imaging and the Gaia RUWE. The red-band campaigns recover the expected depths without the chromatic dependence that would betray an eclipsing binary; the anomalous depths in individual observations are associated with partial or low-quality events (weather-truncated, limited detrending) and not with the signal. Together with the consistent odd/even test and the absence of a secondary eclipse, this rules out the grazing eclipsing-binary scenario that the "V"-shape seen in some light curves might suggest. Additionally, the Gaia DR3 RUWE (1.18) lies below the threshold indicative of multiplicity (1.4; L. Lindegren et al. 2021), consistent with a single star. Since we lack radial velocities, we do not apply the MOLUSC framework (M. L. Wood et al. 2021); the Gemini and SOAR contrast curves incorporated into TRICERATOPS, together with the Gaia RUWE, sufficiently constrain the presence of unresolved companions that could contaminate the signals.

The iterative search, after masking the four objects, reveals an additional signal at $P \approx 28$ days that is not of planetary origin (Figure 3, bottom panel). We note that this systematic reaches a higher SDE (37.3) than the candidate .04 (24.5); the SDE, however, measures detection significance against the noise floor and not planetary nature, so its larger amplitude does not lend it planetary status—unlike candidate .04, whose plausibility rests on transit shape, odd/even consistency, injection–recovery, and dynamical viability rather than on SDE. As the following diagnostics confirm, it shows a discrepant odd/even test ($3.3\sigma$), behavior typical of a detection at half the period and a standard false-positive diagnostic in validation pipelines (N. M. Batalha et al. 2010; J. D. Twicken et al. 2018; J. Li et al. 2019). Moreover, its period coincides closely with twice the TESS orbital period ($2 \times 13.7 = 27.4$ days; the 28.09 day value reflects the finite period resolution and the harmonic-hopping nature of the signal), the scale at which scattered light, momentum dumps, and downlinks inject power into harmonics of the spacecraft's orbital period ($\sim$13.7 days; G. R. Ricker et al. 2015; T. Fetherolf et al. 2023), and when masked it reappears at its harmonic ($\sim$54 days). This pattern of jumping between harmonics is characteristic of a quasi-periodic modulation—of instrumental or stellar-rotation origin (A. McQuillan et al. 2014; R. Angus et al. 2018)—and not of a coherent transit, so we discard it as a planet candidate.

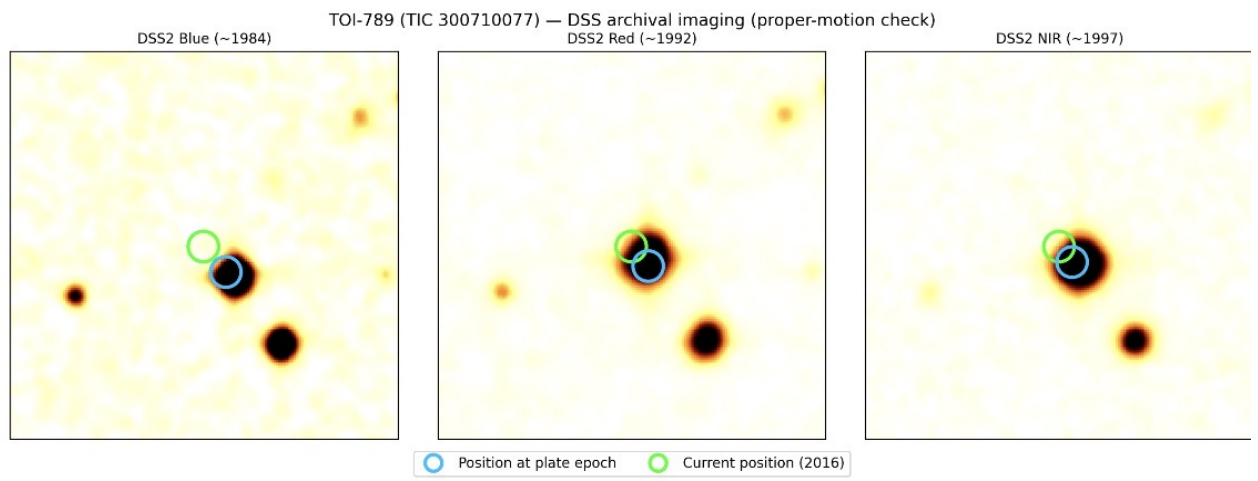

**Figure 5.** Archival imaging from the Digitized Sky Survey (DSS2 Blue $\sim$1984, Red $\sim$1992, and NIR $\sim$1997), used as a proper-motion check. In each panel, the blue circle marks the stellar position at the plate epoch and the green circle the current (2016) position; the large proper motion ($\sim$196 mas yr$^{-1}$) has displaced the star appreciably over the baseline. No background source is seen at the present-day position, excluding a chance-aligned background eclipsing binary as the origin of the signals.

To ensure completeness at long periods, we extended the iterative TLS search to $P = 290$ days, approximately the limit at which the fragmented TESS coverage (78 data blocks over a 2721 day baseline) still yields the $\gtrsim$3 transits required for a reliable period determination (counting only transits falling on epochs with data; C. J. Burke et al. 2014; T. Barclay et al. 2018; S. E. Thompson et al. 2018). No additional planet is recovered. The only power excesses beyond the candidate arise at long periods ($P \approx$ 117–280 days), and several independent lines of evidence identify them as spurious. Most decisively, their significance is highly sensitive to the detrending: under an aggressive median filter (a $\sim$6.7 hr window, far shorter than the stellar-rotation period) the strongest long-period peak collapses from SDE = 28–6.4, as expected for low-frequency stellar or instrumental modulation but not for a genuine, short-duration transit signal. Such sensitivity to the detrending window—and the capacity of detrending to induce or suppress transit-like signals at long periods—is well documented (K. Rodenbeck et al. 2018; M. Hippke & R. Heller 2019). Consistent with this, several of these peaks coincide with low-order multiples of the $\sim$28–29 day stellar-rotation period identified below, indicating residual rotational/activity modulation. This interpretation is further corroborated by the signals themselves: they are supported by too few transits for a reliable period determination: the strongest long-period peak, at $P = 249.7$ days (SDE = 34.7), shows a $6.0\sigma$ odd/even mismatch, while the highest-SDE peaks at $P \gtrsim 250$ days rest on a single transit; others show discrepant (e.g., $3.98\sigma$ at 117 days) or undefined odd/even tests and fitted durations of $\sim$1 hr, far shorter than the $\sim$4.6–6 hr expected for central transits at these periods around this star (whereas candidate .04 shows a duration of 2.2 hr, in line with the 2.6 hr expected). As with the 28 day signal, their large SDE reflects detection significance against the noise floor, not planetary nature. We conclude that candidate .04 ($P = 19.74$ days) is the only robust signal recovered across the full search.

We further tested the origin of this low-frequency signal by computing generalized Lomb–Scargle periodograms (M. Zechmeister & M. Kürster 2009) of both the SAP and PDCSAP light curves after masking all transits (Figure 6).

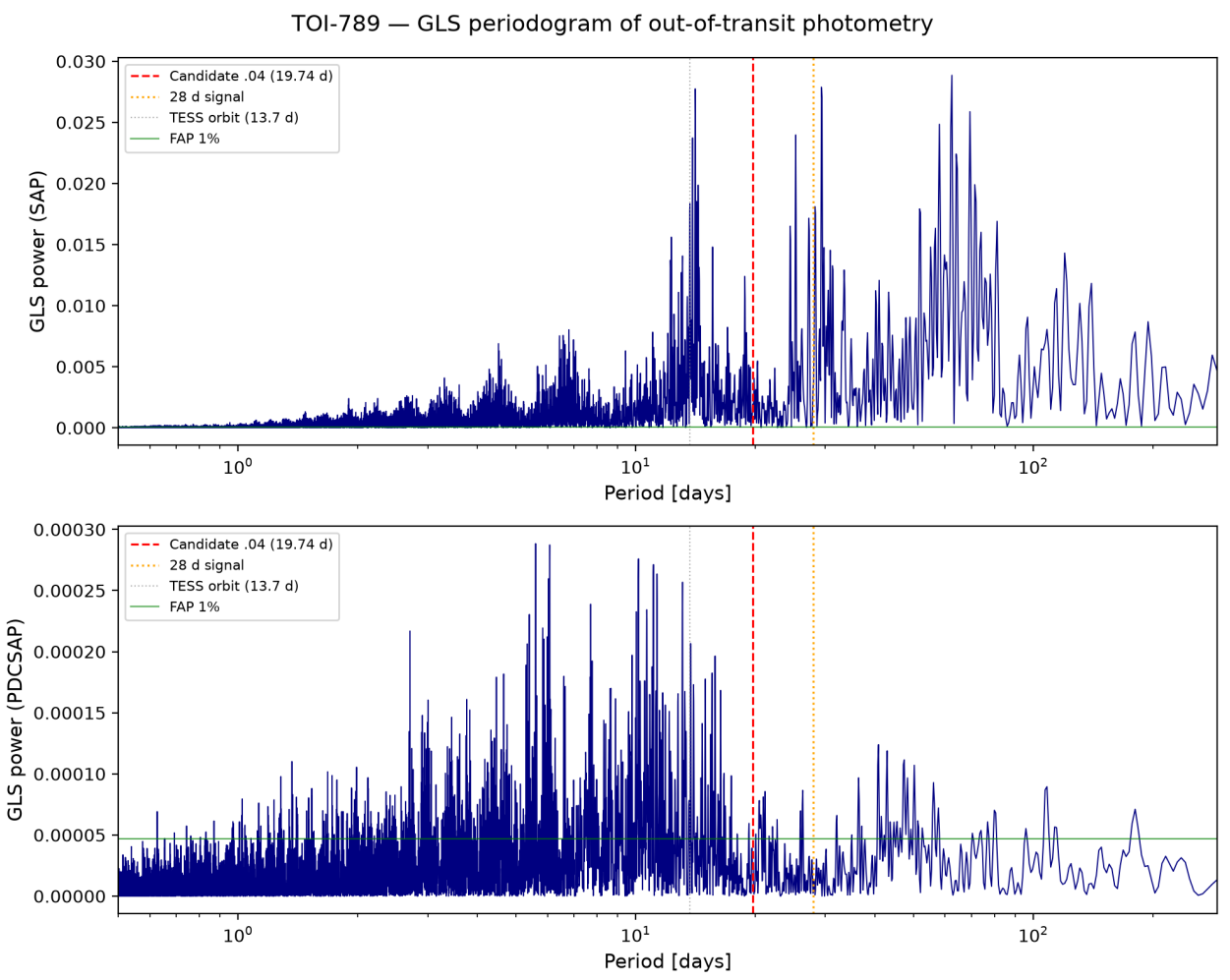

**Figure 6.** Generalized Lomb–Scargle periodograms of the SAP (top) and PDCSAP (bottom) light curves of TOI-789 after masking all transits. The red dashed line marks the candidate's period (19.74 days), the orange dotted line the 28 day signal, and the gray dotted line twice the TESS orbital period (13.7 days); the green line indicates the 1% false-alarm level. Neither light curve shows significant power at 19.74 days. The SAP periodogram is dominated by low-frequency power spanning ∼25–70 days (peaking near 62 days; stellar rotation and its harmonics plus systematics), strongly suppressed in the PDCSAP periodogram, supporting a planetary origin for the candidate.

Crucially, the PDCSAP periodogram shows no significant power at the candidate's period of 19.74 days (below the 1% false-alarm level), nor does it at the 28 day signal or twice the TESS orbital period. The SAP periodogram, which preserves low-frequency stellar variability, is dominated by low-frequency power spanning ∼25–70 days (peaking near 62 days), with additional structure near 27.3 days (consistent with twice the TESS orbital period); we attribute this modulation to a combination of quasi-periodic stellar rotation, harmonics, and instrumental systematics. This power is strongly suppressed in the PDCSAP periodogram, which is expected since the PDC pipeline removes low-frequency stellar variability. The absence of significant GLS power at 19.74 days—in contrast to the strong, transit-shaped TLS signal at that period—supports a planetary rather than a stellar or systematic origin for candidate .04, since a stellar or systematic signal would produce a sinusoidal peak in the periodogram, which is not observed.

## 5. Photometric Analysis

We jointly model the transits with juliet (N. Espinoza et al. 2019), which uses batman (L. Kreidberg 2015) and nested sampling with dynesty (J. S. Speagle 2020). We use the 2 minute cadence SPOC light curve. We flatten the light curve with a moving-median filter masking the transits and trim to windows of ±2.5 durations. We adopt the stellar density as a shared parameter (S. Seager & G. Mallén-Ornelas 2003) with a Gaussian prior from the stellar parameters (Section 3), and a quadratic limb-darkening law (D. M. Kipping 2013). We use the following priors: for each planet, Gaussian priors on the period $P$ and transit time $T_0$ (centered on the TLS values, with widths of 0.001 and 0.02 days, respectively), and uniform priors on the radius ratio $R_p/R_\star \in [0.005, 0.06]$ and the impact parameter $b \in [0, 1]$. The shared stellar density has a Gaussian prior $\rho_\star = 9.9 \pm 1.0$ g cm$^{-3}$ (10% width), and the D. M. Kipping (2013) limb-darkening coefficients have uniform priors $q_1, q_2 \in [0, 1]$. The dilution was fixed to unity in the fit, with the contamination correction (Section 2.1) applied a posteriori to the radii. We include a flux offset (Gaussian prior) and a jitter term (log-uniform prior). The fit recovers a stellar density $\rho_\star = 10.1 \pm 0.9$ g cm$^{-3}$, consistent with the value expected from the stellar parameters (9.9 g cm$^{-3}$), and limb-darkening coefficients $q_1 = 0.69 \pm 0.25$, $q_2 = 0.30 \pm 0.23$ (equivalent to $u_1 = 0.50$, $u_2 = 0.33$). Figure 7 shows the phase-folded transits with the best-fit model and their residuals.

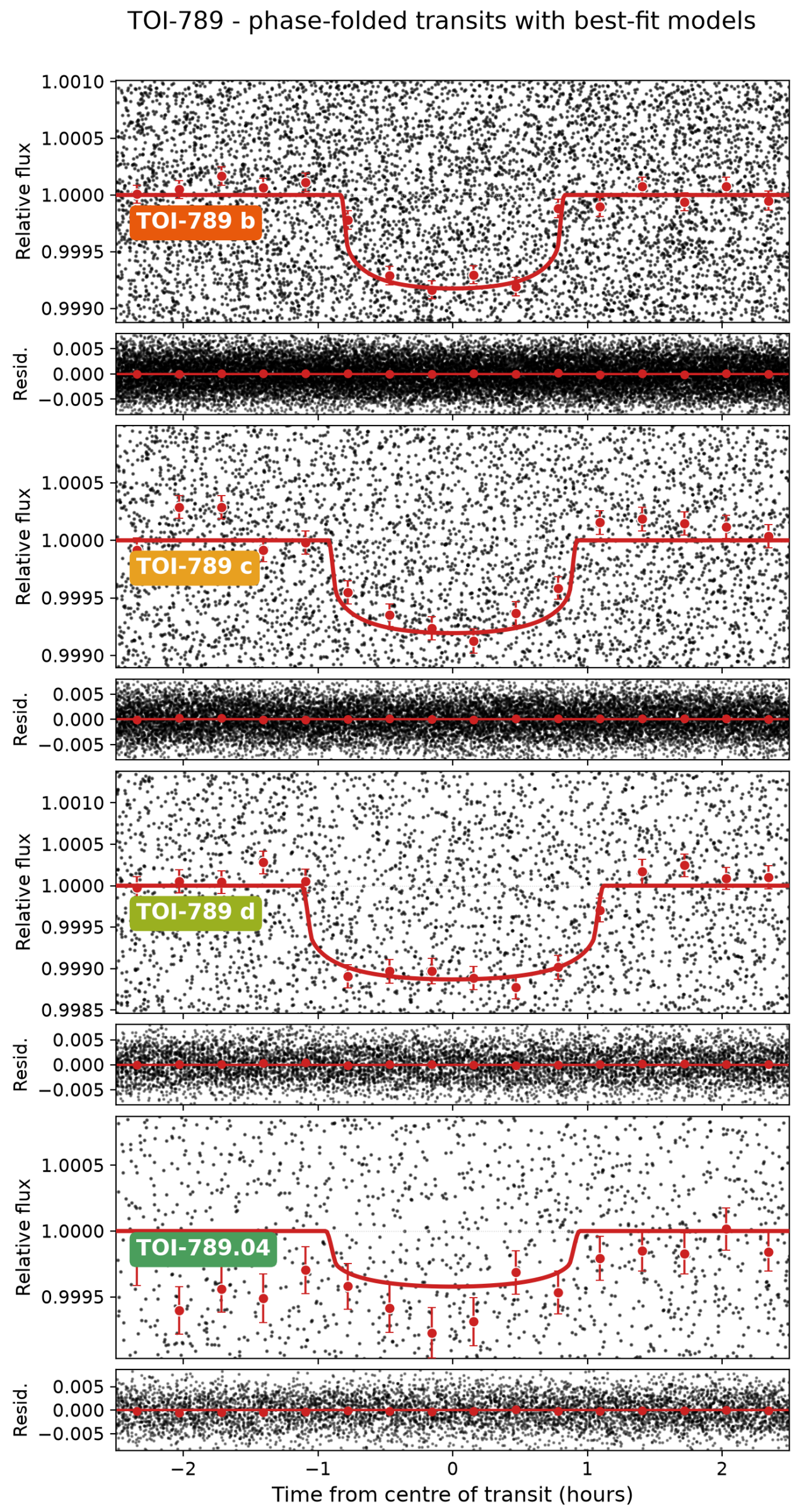

**Figure 7.** Phase-folded TESS transits of TOI-789 b, c, d, and the candidate .04 (top to bottom), each folded on its best-fit period. Gray points are the unbinned PDCSAP data; colored points are phase-binned for clarity. The red curve is the best-fit joint juliet model (Section 5); the lower strip of each panel shows the residuals, consistent with photometric noise. The shallow depths (∼1 ppt) and the decreasing transit signal-to-noise from b to the candidate .04 are apparent.

We fit circular orbits ($e = 0$). We compare the Bayesian evidence of a circular model against one with free eccentricity,

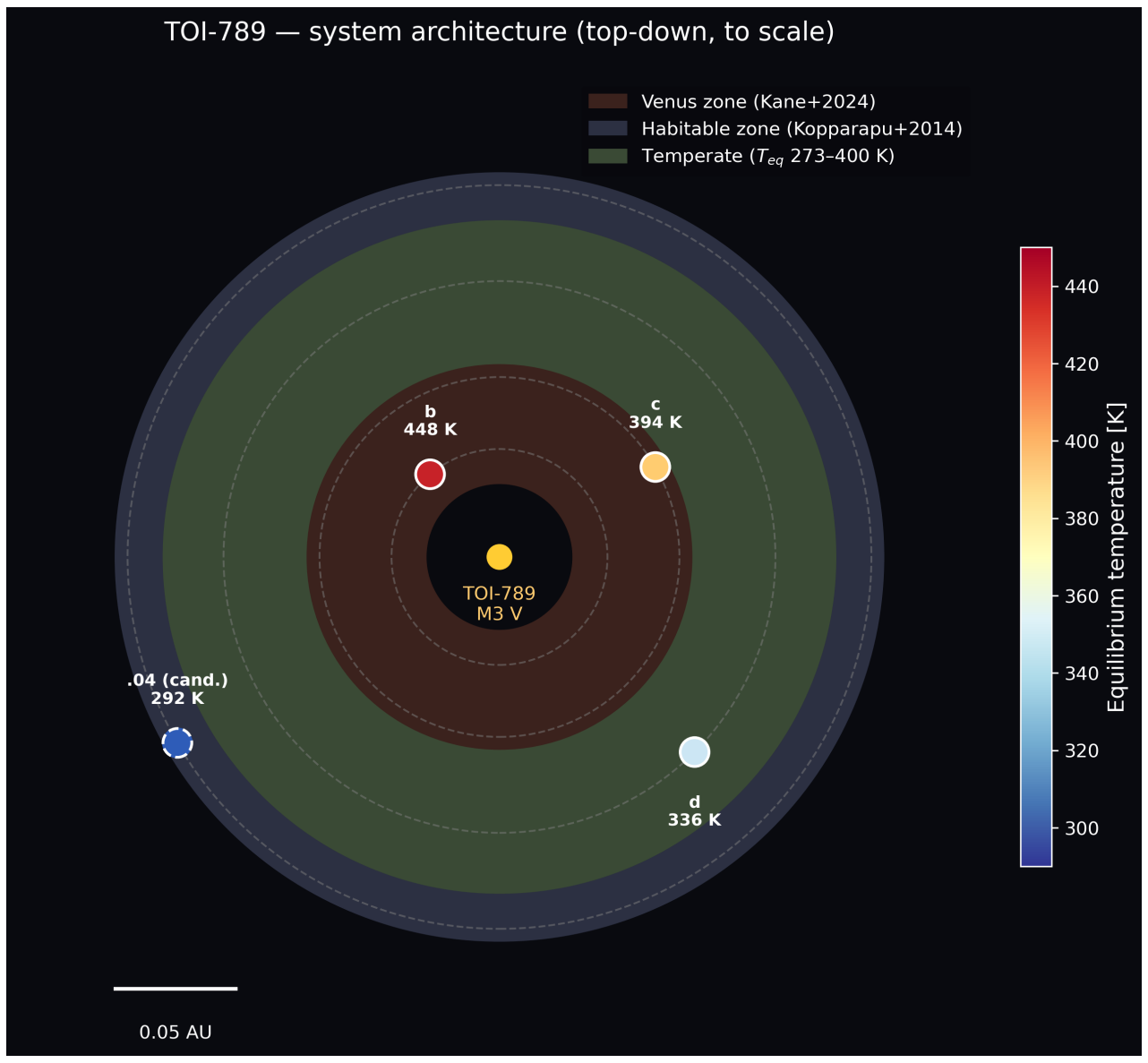

**Figure 8.** Top-down, to-scale view of the TOI-789 architecture. The four planetary orbits are drawn around the M3 host (center); symbol size scales with planetary radius and symbol color with equilibrium temperature (color bar). The candidate .04 is drawn with a dashed outline. Shaded annular bands mark, from the inside out, the Venus zone (S. R. Kane et al. 2014), the temperate region defined by $T_{\rm eq} = 273$–400 K, and the inner edge of the R. K. Kopparapu et al. (2014) habitable zone. The four planets form a sequence of decreasing insolation that crosses the Venus zone, with the candidate .04 near the inner edge of the habitable zone.

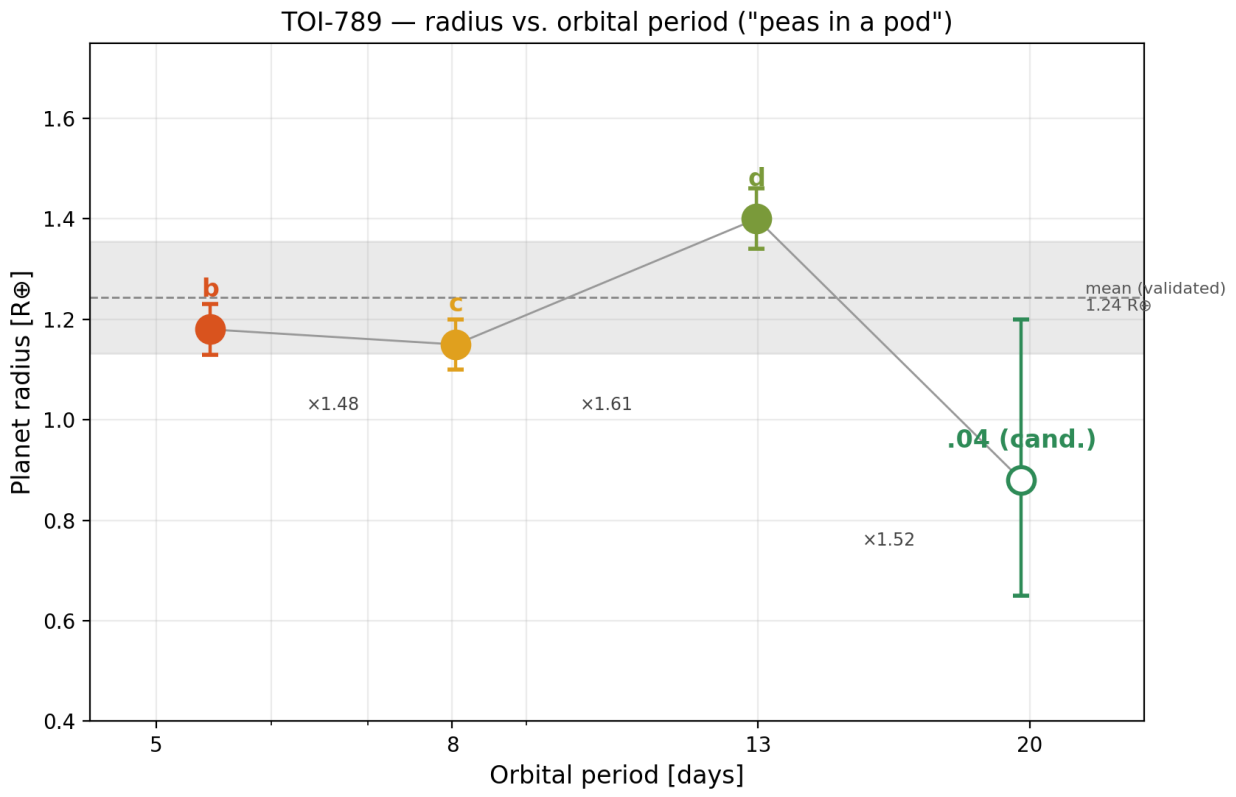

**Figure 9.** Planetary radius vs. orbital period for TOI-789 ("peas in a pod" diagram). Validated planets (b, c, d) are shown as filled symbols and the candidate .04 as an open symbol, with $1\sigma$ uncertainties (the candidate's asymmetric error reflects the radius–impact-parameter degeneracy). The dashed line and gray band mark the mean radius of the validated planets and its $\pm 1\sigma$ scatter; the multiplicative period ratios between adjacent planets are annotated. The radii are uniform to within the scatter—d being the mild outlier—and the spacing is regular, both signatures of the intrasystem uniformity discussed in Sections 6.1 and 8.1.

obtaining $\Delta \ln Z = -1.0$, which indicates no significant preference for the eccentric model (R. Trotta 2008); we therefore adopt the more parsimonious circular model. Having adopted the circular model, we do not report eccentricity limits from the free-eccentricity fit. The constraint on the eccentricities comes from the dynamical stability analysis (Section 6.2), which limits $e \lesssim 0.05$.

After applying the dilution correction (+8.2%; Section 2.1), we adopt as definitive planetary radii $R_{p,b} = 1.18 \pm 0.05$, $R_{p,c} = 1.15 \pm 0.05$, and $R_{p,d} = 1.40 \pm 0.06\,R_\oplus$ (terrestrial regime), and $R_{p,.04} = 0.88$ (0.65–1.20) $R_\oplus$ for the candidate. These values are consistent with the on-target depths recovered from the ground in the red band in uncontaminated apertures (Section 2.2), which independently confirm the dilution of the TESS photometry. The full parameters are listed in Table 3. We note that the period-order nomenclature (b, c, d) corresponds to TOI-789.01, .03, .02 respectively.

## 6. Architecture and Dynamical Stability

### 6.1. Orbital Configuration

The planets form a compact system with periods of 5.45, 8.04, 12.97 (and 19.74) d (Figure 8). The period ratios between adjacent planets ($P_c/P_b = 1.48$, $P_d/P_c = 1.61$, $P_{.04}/P_d = 1.52$) lie close to—but not at—low-order commensurabilities (3:2, 8:5, 3:2); the resonant angles circulate rather than librate, confirming that the system is near-resonant but not resonant. The dynamical spacing is $\Delta \approx 14$ mutual Hill radii (versus the typical Kepler value of ∼20; L. M. Weiss et al. 2018), placing TOI-789 among the compact systems. The uniform radii and regular spacing are consistent with the "peas in a pod" pattern (Figure 9; S. Millholland et al. 2017; L. M. Weiss et al. 2018).

### 6.2. Dynamical Stability

We assess stability with SPOCK (D. Tamayo et al. 2020) and $N$-body integrations with REBOUND/WHFast (H. Rein & S.-F. Liu 2012; H. Rein & D. Tamayo 2015), with masses from the J. Chen & D. Kipping (2017) mass–radius relation, following an approach analogous to recent dynamical analyses of compact TESS multiplanet systems (e.g., M. Greklek-McKeon et al. 2026).

Our primary analysis considers the three validated planets (b, c, d), since TOI-789.04 remains a candidate. For this configuration, direct $N$-body integrations ($10^7$ orbits, three realizations per configuration) indicate robust stability at low eccentricity: configurations with $e \leqslant 0.05$ are stable in all realizations (3/3), while at $e = 0.10$ the system mostly destabilizes (1/3 stable) and at $e = 0.15$ entirely (0/3). SPOCK indicates a robustly stable configuration ($P \approx 0.97$) up to $e \approx 0.15$, degrading only for $e \gtrsim 0.20$. A mass sweep at low eccentricity shows stability up to a factor of ∼5 above the nominal masses. The mass–eccentricity map (Figure 10) reveals a coupling: the stability limit goes from $M_d \lesssim 16\,M_\oplus$ at $e \lesssim 0.07$ to $M_d \lesssim 8\,M_\oplus$ at $e \approx 0.13$. We conclude that the three-planet validated system is dynamically stable for $e \lesssim 0.05$, with eccentricity—not mass—being the limiting factor.

As a complementary analysis, we evaluate the conditional scenario in which the candidate .04 is real (four-planet system), incorporating it as a viability test: configurations that lead to rapid instability relative to the system's age can be rejected as unphysical, allowing the eccentricities to be bounded even in the absence of measured masses (J. J. Lissauer et al. 2011; D. Tamayo et al. 2020). In this case, the same eccentricity threshold holds: configurations with $e \leqslant 0.05$ remain stable. Direct $N$-body integrations (REBOUND/WHFast, $10^7$ orbits, five realizations per configuration) reveal a sharp stability threshold: configurations with $e \leqslant 0.05$ are stable in all realizations (5/5), whereas at $e = 0.10$ the system destabilizes in four of five cases—in fewer than $5 \times 10^5$ orbits—and at $e = 0.15$ in all of them. SPOCK, trained to predict stability over longer timescales ($10^9$ orbits), yields a somewhat stricter constraint ($P = 0.92$, 0.42, and 0.17 at $e = 0.02$, 0.05, and 0.10 respectively), a difference consistent with the steep growth of the instability time with orbital separation (A. Obertas et al. 2017) and with long-term instability not captured by the $10^7$-orbit integrations. The two methods agree on the qualitative trend—that eccentricity, rather than mass, sets the stability limit and that tolerance to eccentricity drops sharply once the fourth planet is included—but differ quantitatively at the boundary: the $N$-body integrations admit $e = 0.05$ (5/5 stable), whereas SPOCK, which probes $10^9$ orbits, already disfavors it ($P = 0.42$) and points to a stricter limit ($e \lesssim 0.02$–0.05). Taken together they imply that the presence of the fourth planet requires low eccentricities ($e \lesssim 0.05$) for the system's survival, reinforcing—if the candidate is confirmed—the dynamically cold nature of the system. The stability of both scenarios is summarized in Table 4.

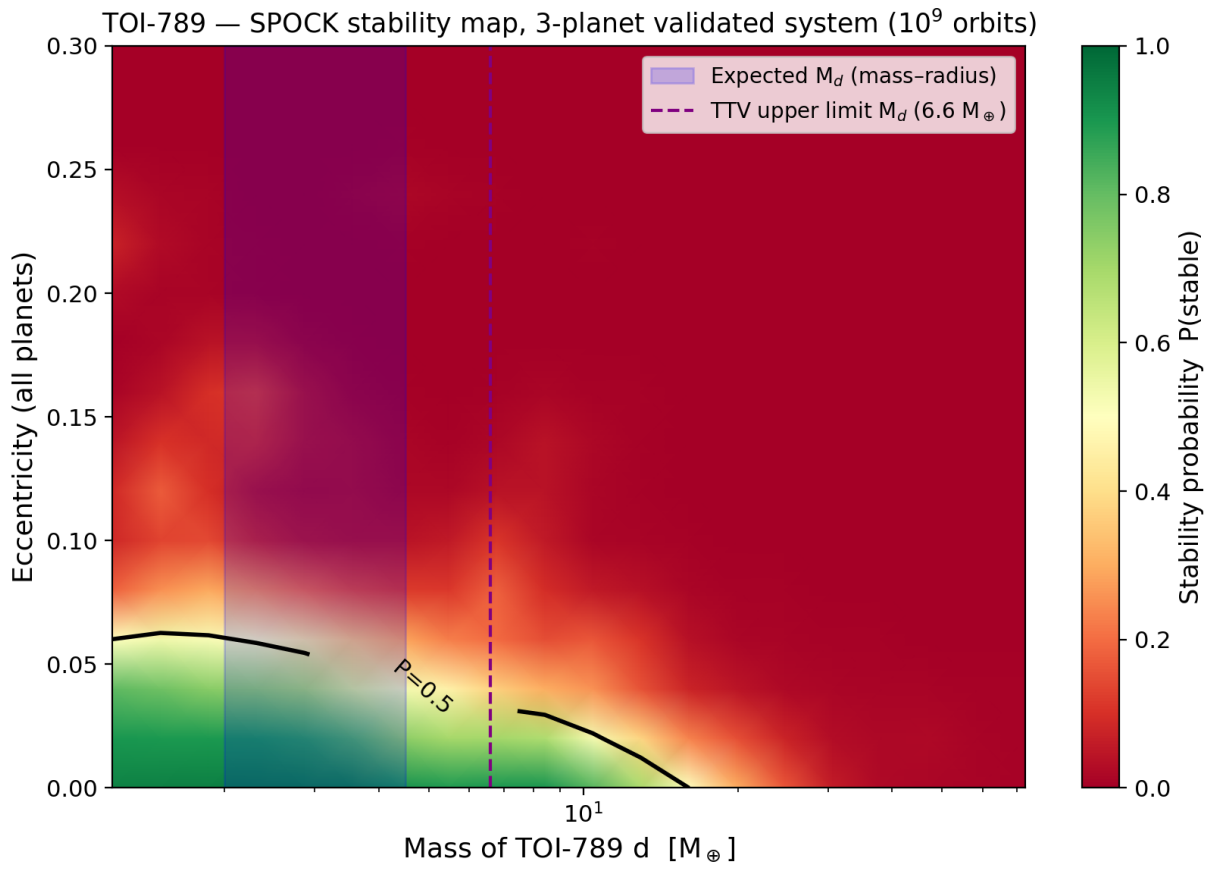

**Figure 10.** SPOCK dynamical stability map for the three-planet validated system, computed over $10^9$ orbits, in the plane of TOI-789 d mass vs. the common orbital eccentricity of all planets, with the masses of b and c held fixed at their nominal mass–radius values (J. Chen & D. Kipping 2017); d is varied because it is the most massive, dynamically dominant body. Color encodes the stability probability $P$(stable) and the black contour marks $P$ (stable) = 0.5. The shaded vertical band indicates the mass of d expected from its measured radius (1.40 $R_\oplus$). The map shows that stability requires low eccentricity ($e \lesssim 0.05$ at the expected mass), eccentricity rather than mass being the limiting factor.

Survival on gigayear timescales requires low eccentricities (J. Fang & J.-L. Margot 2013; B. Pu & Y. Wu 2015), increasingly so the higher the true multiplicity, since the instability time of a compact system decreases as the number of planets increases at fixed separation (J. E. Chambers et al. 1996; A. W. Smith & J. J. Lissauer 2009). Combined with the compact spacing and uniform radii, this indicates that TOI-789 is a dynamically cold system, consistent with a quiescent formation history.

### 6.3. Transit-timing Variations and Mass Limits

We extracted individual transit times for the three validated planets from the TESS photometry (171, 114, and 62 transits for b, c, and d, respectively) and constructed the $O - C$ diagram relative to a linear ephemeris (Figure 11). The transit times show no significant deviations from a linear ephemeris (per-planet rms of 8.5, 11.2, and 10.4 minutes, comparable to the individual timing uncertainties), with no trends or periodicities: we detect no significant TTVs, as expected for a configuration far from first-order mean-motion resonances, in whose vicinity the TTV signal is amplified (E. Agol et al. 2005; M. J. Holman & N. W. Murray 2005; Y. Lithwick et al. 2012). The nondetection of TTVs nonetheless yields upper limits on the masses via a TTVFast (K. M. Deck et al. 2014) analysis, with the masses (in log space) and eccentricities as free parameters sampled with

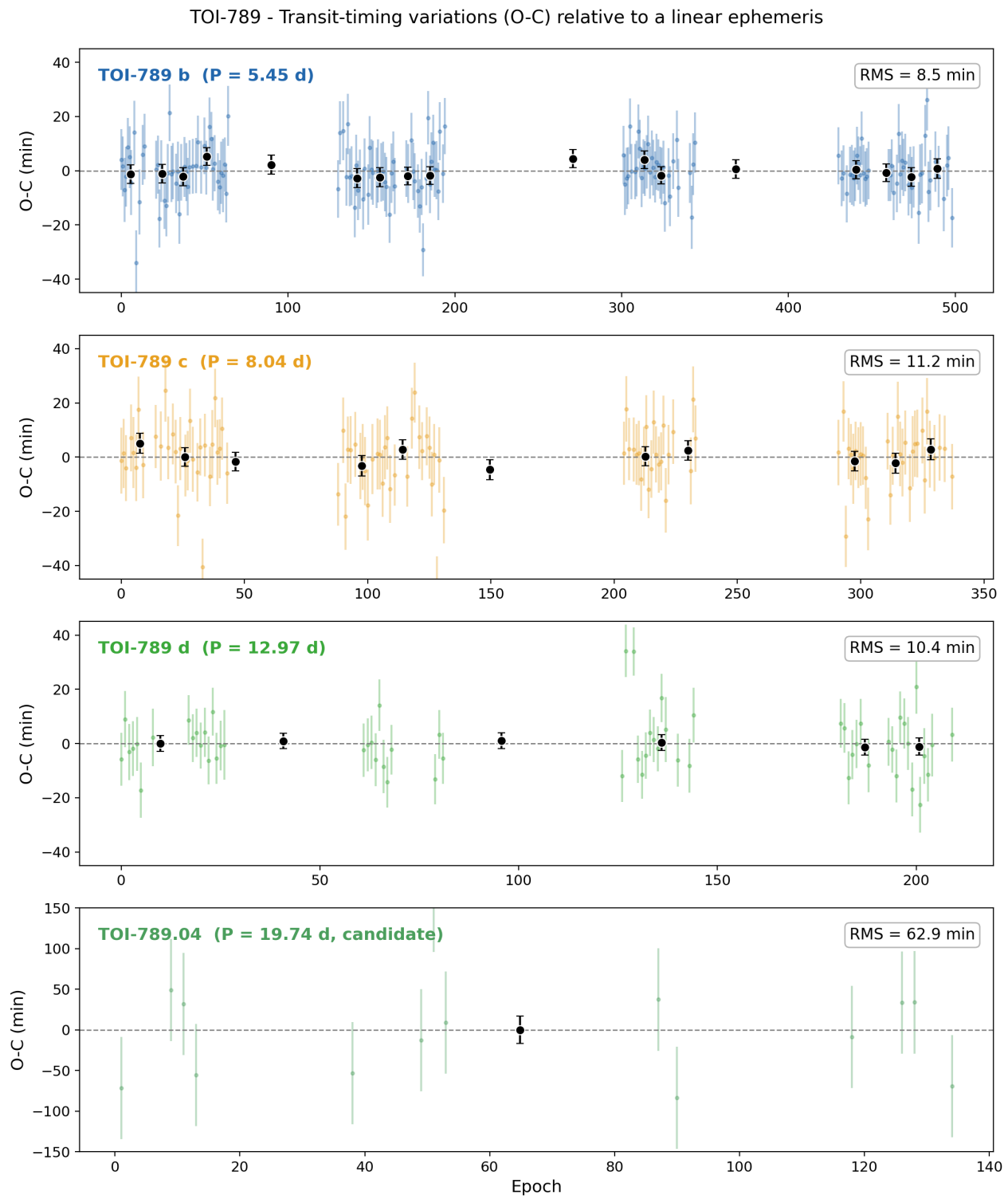

**Figure 11.** Transit-timing variations ($O - C$) of TOI-789 b, c, d, and the candidate .04 relative to a linear ephemeris, derived from the TESS photometry (171, 114, 62, and 14 transits, respectively). Each point is an individual transit time with its 1$\sigma$ uncertainty; black points are epoch-binned averages. The per-planet rms values (8.5, 11.2, 10.4, and 62.9 minutes) are comparable to the individual timing uncertainties for the validated planets and show no significant trend or periodicity, consistent with linear ephemerides and with the absence of detectable TTVs expected for a configuration away from first-order mean-motion resonances. The candidate .04 shows larger scatter, reflecting its lower SNR.

**Table 4**
Dynamical Stability

| Ecc. | $N$-body ($10^7$ orb.) | SPOCK ($10^9$ orb.) |
|---|---|---|
| | Three-planet system (b, c, d) | |
| 0.02 | 3/3 stable | Robustly stable |
| 0.05 | 3/3 stable | Stable |
| 0.10 | 1/3 stable | Marginal |
| 0.15 | 0/3 stable | Unstable |
| | Four-planet system (incl. .04) | |
| 0.02 | 5/5 stable | $P = 0.92$ |
| 0.05 | 5/5 stable | $P = 0.42$ |
| 0.10 | 1/5 stable | $P = 0.17$ |
| 0.15 | 0/5 stable | … |

**Note.** Direct $N$-body integrations (REBOUND/WHFast, five realizations) and the SPOCK classifier (D. Tamayo et al. 2020). The differing thresholds reflect the distinct timescales probed ($10^7$ vs. $10^9$ orbits). The two methods agree on the trend—eccentricity limits stability and the fourth candidate requires $e \lesssim 0.05$ to survive—though SPOCK, probing longer timescales, is the more conservative, already disfavoring $e = 0.05$ ($P = 0.42$).

emcee (D. Foreman-Mackey et al. 2013): we obtain $M_b < 3.7$, $M_c < 3.0$, and $M_d < 6.6\,M_\oplus$ (95% confidence), consistent with the terrestrial regime and with the independent dynamical upper limits (Section 6.2). A measurement of individual masses, however, remains out of reach.

We repeated the analysis including the candidate TOI-789.04 as a fourth body. Its $O - C$ diagram, with 14 measurable transit times (versus 62–171 for the validated planets, reflecting its lower SNR and a larger $O - C$ scatter of $\sim$63 minutes), is likewise consistent with a linear ephemeris. The four-body fit yields an upper limit of $M_{.04} < 2.1\,M_\oplus$, consistent with the expected mass for a terrestrial body of its radius, without altering the limits of the validated planets. The inclusion of the candidate confirms the absence of detectable TTVs across the system.

## 7. Temperate Zone, Habitability, and Characterization

### 7.1. The Concept of the Temperate Zone

The notion of a “temperate” planet lacks a unique definition. Proposals are split between insolation criteria—0.5–1.5 $S_\oplus$ (N. B. Cowan et al. 2015), 0.25–4 $S_\oplus$ (A. H. M. J. Triaud et al. 2024)—and equilibrium-temperature criteria, such as the 273–395 K interval of M. N. Günther et al. (2019), bounded by the freezing point of water and the survival temperature of terrestrial extremophiles (we adopt a slightly rounded upper cap of $T_{\rm eq} \leqslant 400$ K throughout, see Section 7.2). M. G. Scott et al. (2026) propose a definition that combines both, setting the upper insolation limit (5 $S_\oplus$) at the 75th percentile of the distribution of known temperate planets while retaining a temperature cap ($T_{\rm eq} \leqslant 400$ K). We preferentially adopt the equilibrium-temperature criterion. This choice rests on a methodological consideration: a threshold calibrated on the distribution of the observed sample inevitably reflects the selection functions of transit surveys—which favor short-period, higher-insolation planets—and evolves as the census grows and becomes complete. A criterion anchored to physically meaningful temperatures remains invariant against such biases. Both perspectives are complementary—and converge on the $T_{\rm eq} \leqslant 400$ K cap; here we favor the physically grounded one for its greater conceptual stability.

### 7.2. Insolation and the Sequence Across the Venus Zone

The insolation and equilibrium temperature of the four objects are listed in Table 3.

The equilibrium temperatures are computed assuming a Bond albedo of 0.3 and full heat redistribution, following the TESS Input Catalog convention (K. G. Stassun et al. 2019); with zero albedo, the values would be about 10% higher. The four objects, stepped in insolation (1.71–9.5 $S_\oplus$) around a single star—and therefore with identical age, metallicity, and irradiation history—span the range in which a terrestrial atmosphere would transition to the runaway-greenhouse state (Venus zone; S. R. Kane et al. 2014). This configuration makes them, in principle, a case study for investigating the evolutionary divergence between Earth-like and Venus-like planets. Under the adopted equilibrium-temperature criterion ($T_{\rm eq} \leqslant 400$ K), three of the four objects fall in the temperate regime: c (394 K) and d (336 K), both validated, plus the candidate .04 (292 K), the last near the optimistic (recent-Venus) inner edge of the habitable zone ($S_{\rm inner} \approx 1.6\,S_\oplus$; R. K. Kopparapu et al. 2014). Their high insolations (5.7 and 3.0 $S_\oplus$ for c and d), however, place them on the Venus side of the transition, making the system a laboratory of the divergence itself.

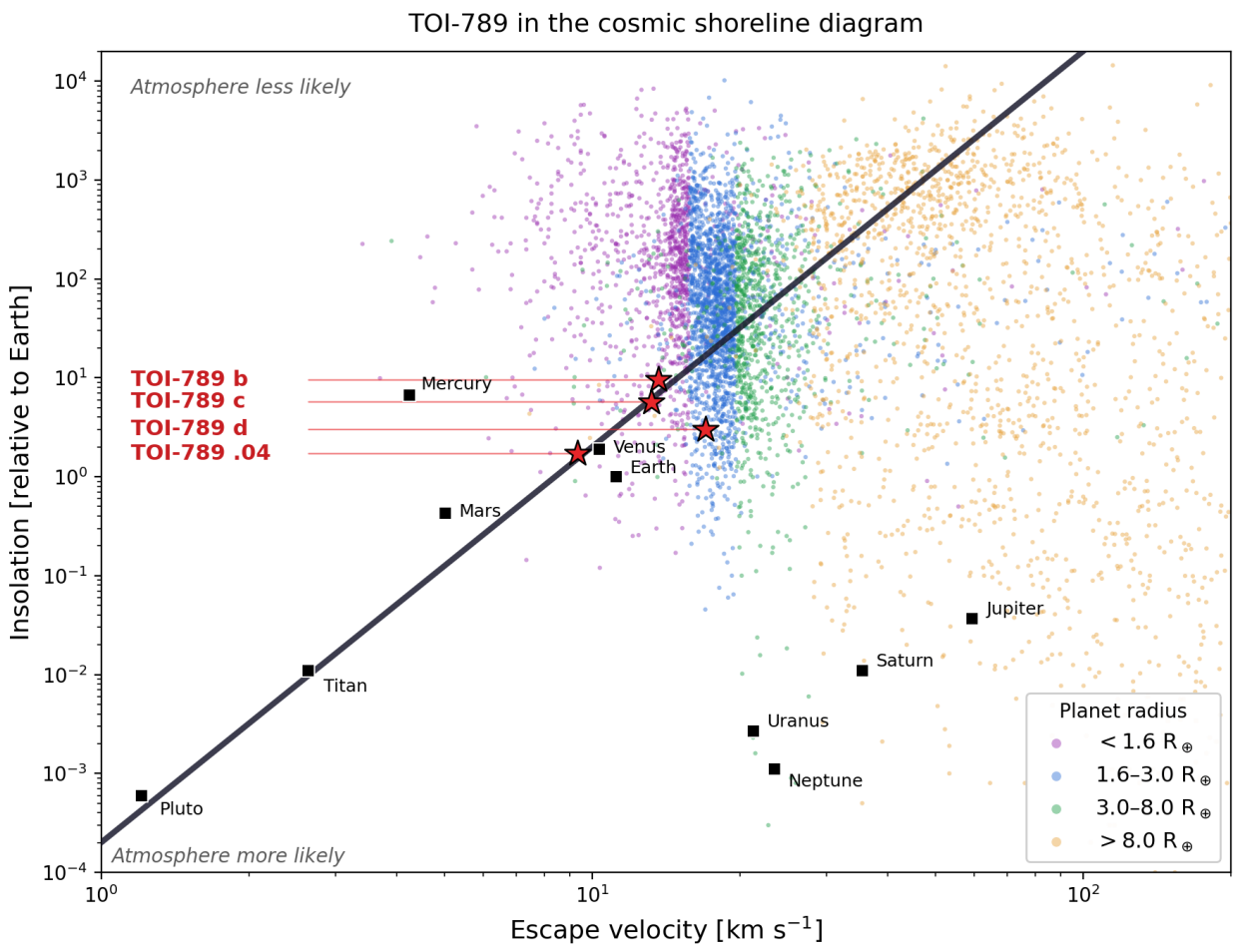

**Figure 12.** Location of the four TOI-789 planets in the cosmic shoreline diagram (K. J. Zahnle & D. C. Catling 2017): bolometric insolation (relative to Earth) vs. escape velocity. The planets are shown as stars colored per planet; escape velocities are computed from the dilution-corrected radii via the J. Chen & D. Kipping (2017) mass–radius relation. The background population is the known exoplanet sample (NASA Exoplanet Archive), colored by radius bin, and solar system bodies are shown as black squares. The diagonal line is the empirical shoreline separating bodies more likely to retain an atmosphere (lower right) from those less likely to (upper left). The innermost planet b lies in the loss regime despite a moderate escape velocity, owing to its high insolation; d lies most clearly on the retention side. The diagram uses bolometric insolation and does not account for the enhanced XUV history of M dwarfs (Section 7.3).

### 7.3. Atmospheric Retention and Characterization Prospects

We place the four objects of the system in the cosmic shoreline plane (K. J. Zahnle & D. C. Catling 2017; Figure 12), using quantities derived directly from our data: the insolation and the escape velocity (the latter from the dilution-corrected radii and the J. Chen & D. Kipping 2017 mass–radius relation). The result illustrates the interplay between irradiation and gravity: the innermost planet, b, despite its moderate escape velocity ($v_{\rm esc} \sim 13.7$ km s$^{-1}$), lies in the atmospheric-loss regime owing to its high insolation (9.5 $S_\oplus$), being the most vulnerable in the system; d, with the largest radius and escape velocity ($\sim$17 km s$^{-1}$) and moderate insolation, lies most clearly in the retention regime; c is near the boundary; and the fourth candidate, .04, lies in the loss regime but less deeply than b, owing to its low insolation (1.71 $S_\oplus$).

For the candidate .04 in particular, the retention/loss threshold is reached around $R_p \sim 0.9\,R_\oplus$: once the radii are corrected for dilution, an appreciable fraction of its allowed radius range (0.65–1.2 $R_\oplus$) corresponds to the retention regime. Its position is therefore ambiguous and sensitive to the radius.

However, a caveat inherent to the stellar type should be emphasized: the Zahnle & Catling formulation in terms of bolometric insolation does not explicitly account for the integrated XUV flux, which in M dwarfs is disproportionately high and persists throughout their extended activity phase. As E. K. Pass et al. (2025) show, incorporating this flux shifts the cosmic shoreline toward the loss regime for M-dwarf planets.

More broadly, hydrodynamic-escape treatments that move beyond the energy-limited prescription find that, once a range of initial volatile inventories and atmospheric compositions is allowed, the shoreline is better described as a broad transition zone than as a sharp boundary (X. Ji et al. 2025), further blurring the retention/loss assignment for marginal, low-mass objects such as candidate .04 (which fall below the super-Earth regime where atomic-line cooling can render atmospheres resilient). The results obtained here in the insolation plane should therefore be contrasted with these XUV- and escape-physics-based formalisms (X. Ji et al. 2025; E. K. Pass et al. 2025) to reach a firm conclusion, an analysis that requires constraining the rotational history of the star, which is not reliably determined by current data (Section 6).

Regardless of the atmospheric-retention question, the system's characterization prospects are limited: the transmission spectroscopy metric (TSM $\sim$ 3–5; E. M.-R. Kempton et al. 2018), together with the moderate brightness of the host ($T_{\rm mag} \sim 12$) and its distance (43 pc), place TOI-789 beyond the reach of JWST. The exploitation of its potential as a comparative-planetology laboratory across the Venus zone is therefore contingent on future facilities with greater collecting power.

## 8. Discussion

### 8.1. TOI-789 in the Context of Compact M-dwarf Systems

TOI-789 adds to the growing population of compact multiplanet systems around M dwarfs characterized by TESS, an example of the "peas in a pod" pattern in its components' size uniformity and regular spacing. The large radius uncertainty of the outer candidate (.04: 0.65–1.20 $R_\oplus$, dominated by the radius–impact-parameter degeneracy) precludes evaluating the size-gradient component: although its nominal value is the smallest of the four objects, within the error bars it is compatible with the radii of the inner terrestrial planets. Among the three validated planets, the radii are uniform, except that of d, which is slightly larger ($\sim 3\sigma$ above b and c; L. M. Weiss et al. 2018). The all-terrestrial architecture of TOI-789, with no sub-Neptune among its members, is consistent with the disappearance of the radius valley around mid-to-late M dwarfs, where the close-in small-planet population becomes unimodal and dominated by super-Earths (K. Ment & D. Charbonneau 2023; E. D. Gillis et al. 2026). All four objects ($R_p = 1.18$, 1.15, 1.40, and 0.88 $R_\oplus$) lie below the M-dwarf radius valley, located at $1.64 \pm 0.03\, R_\oplus$ (H. M. Parashivamurthy & G. D. Mulders 2025), placing the entire system on the super-Earth/rocky side of the gap and reinforcing its terrestrial interpretation. Its closest analog is L 98-59 (V. B. Kostov et al. 2019), a system of terrestrial planets around an M3 dwarf; however, L 98-59 (10.6 pc, bright host) has allowed RV masses, additional planets, and JWST characterization, whereas TOI-789 (43 pc) represents the—numerically dominant—case whose detailed characterization is limited by host brightness. Other analogs include TOI-270 (M. N. Günther et al. 2019), K2-32 (R. Heller et al. 2019, also with a small fourth planet found by iterative TLS), and especially HD 101581 (M. Kunimoto et al. 2025). The latter bears a close methodological and architectural parallel to TOI-789: both are compact, peas-in-a-pod, near-resonant (without libration), statistically validated systems, and both host an additional planet not validated owing to its low signal-to-noise ratio. However, they differ in two relevant respects. First, in observability, at $V = 7.77$, HD 101581 is the brightest known star hosting multiple subterrestrial planets, fully accessible to characterization, whereas TOI-789—4 times more distant and considerably fainter—represents the opposite extreme: not characterizable. Second, in insolation regime, the planets of HD 101581 are all too hot ($T_{\rm eq}$ = 690–834 K) to probe the transition—they are already at the post-greenhouse extreme—whereas TOI-789 hosts planets stepped on both sides of the runaway-greenhouse threshold, including two validated planets in the temperate regime under the adopted criterion (c, 394 K; d, 336 K) and the candidate .04 (292 K) near the inner edge of the habitable zone. This sequence—anchored on confirmed planets and not dependent on the confirmation of .04—makes TOI-789 a potentially more informative laboratory than HD 101581 for studying the Earth–Venus evolutionary divergence around a single star, once instrumentation allows (Section 7.3). HD 101581 retains, in contrast, the decisive advantage of immediate observability; TOI-789 thus occupies a complementary niche, of greater scientific interest but with characterization contingent on future facilities.

### 8.2. The Near-resonant Architecture

Period ratios close to low-order commensurabilities without resonant trapping are characteristic of the Kepler population (J. J. Lissauer et al. 2011; D. C. Fabrycky et al. 2014), interpreted as convergent migration followed by departure from resonance. Unlike resonant chains such as TRAPPIST-1 (M. Gillon et al. 2017) or HD 110067 (R. Luque et al. 2023), TOI-789 represents the more common nonresonant configuration.

### 8.3. The Fourth Candidate and Dynamical Coherence

The detection of the candidate .04, although not validatable with current data, is supported by two independent arguments: it is recoverable in injection–recovery tests (J. L. Christiansen et al. 2015; R. Heller et al. 2019) and it is dynamically viable. This dynamical coherence—its presence does not introduce unstable configurations and restricts the eccentricities to $e \lesssim 0.05$, in keeping with the dynamically cold and packed nature of the system (J. Fang & J.-L. Margot 2013; B. Pu & Y. Wu 2015; D. Tamayo et al. 2020, 2021)—reinforces, without confirming, its planetary plausibility. Its confirmation will require higher-precision photometry.

### 8.4. Follow-up Prospects

Since the cumulative SNR grows roughly as the square root of the number of transits, reaching the validation threshold (SNR $\gtrsim 10$) would require approximately doubling the number of TESS transits observed; alternatively, a small number of ground-based transits with higher photometric precision would provide a much higher per-transit SNR and could suffice for confirmation. Confirmation of the candidate requires higher-precision transit photometry—e.g., LCOGT (T. M. Brown et al. 2013), MEarth (P. Nutzman & D. Charbonneau 2008; J. M. Irwin et al. 2015), or SPECULOOS (L. Delrez et al. 2018; D. Sebastian et al. 2021)—or reobservation in future TESS sectors. RV mass measurement is infeasible ($K \lesssim 1$ m s$^{-1}$ for $T_{\rm mag} \approx 12$). The ground-based follow-up (ExoFOP-TESS) shows no significant TTVs: the reported

**Table 5**
Predicted Transit Ephemerides (Example Windows, 2026)

| Planet | Oct 2026 Mid-transit Times (UTC) | $\sigma$ |
|---|---|---|
| b | Oct 6, 11, 16, 22, 27 | ±4 minutes |
| d | Oct 8, 16, 24 | ±8 minutes |
| c | Oct 4, 17, 30 | ±6 minutes |
| .04 (cand.) | Oct 10 (22:47), Oct 30 (16:37) | ±35 minutes |

**Note.** Predicted from the refined ephemerides; uncertainties propagated as $\sqrt{\sigma_{T_0}^2 + (n\ \sigma_P)^2}$. The candidate .04 has a substantially larger uncertainty owing to its poorly constrained ephemeris.

deviations in the transit times ($\lesssim$20 min, at $1\sigma$–$1.5\sigma$) are compatible with linear ephemerides (see Section 6.3).

Dedicated spectroscopy would allow refining the metallicity via near-infrared calibrations (B. Rojas-Ayala et al. 2012; A. W. Mann et al. 2013) and, with it, the stellar radius and the planetary radii (A. W. Mann et al. 2015, 2019). In the longer term, high-resolution spectroscopy with 30 m class telescopes could open access to the atmospheric characterization of temperate planets around nearby M dwarfs (I. A. G. Snellen et al. 2013; F. Rodler & M. López-Morales 2018; M. López-Morales et al. 2019), although the distance (43 pc) and moderate brightness of TOI-789 would place it among the most demanding targets even for those facilities.

Regardless of those future prospects, the immediate value of this work is operational and methodological. The ephemerides refined over 42 sectors allow scheduling transit windows for the candidate .04 with low uncertainty, so that its confirmation or rejection is within reach of the ground-based follow-up networks (LCOGT, MEarth, SPECULOOS) and the upcoming sectors of the TESS extended mission, without prior reanalysis. Table 5 lists example predicted transit windows for the candidate .04 during 2026 to facilitate such follow-up. The analysis chain employed—iterative TLS search with masking, injection–recovery, validation with TRICERATOPS (with resolved-neighbor removal informed by the ground-based photometry), and dynamical constraint with $N$-body and SPOCK—constitutes a reproducible template for other compact systems around faint M dwarfs in which individual validation falls short. Likewise, the system, dynamically characterized, is incorporated as an additional case into the samples of near-resonant and packed architectures relevant for population studies (J. Fang & J.-L. Margot 2013; B. Pu & Y. Wu 2015; L. M. Weiss et al. 2018), and the treatment of the ~28 day systematic signal and of the dilution by the in-pixel contaminant offer practical, reusable cautions for the analysis of multisector light curves of contaminated targets.

## 9. Conclusions

1. We statistically validate three planets (TOI-789 b, c, d) with TRICERATOPS; after excluding the resolved neighbors ruled out by the on-target ground-based photometry, all three satisfy both validation criteria (FPP $< 0.015$ and NFPP $< 10^{-3}$), with the multiplicity prior providing additional reinforcement. They are terrestrial planets with uniform radii (1.18, 1.15, 1.40 $R_\oplus$) and periods of 5.45, 8.04, and 12.97 days.
2. We detect a fourth candidate (TOI-789.04, $P \approx 19.74$ days, $R_p \approx 0.88\,R_\oplus$) via iterative TLS, not previously cataloged. Its low SNR precludes formal validation; we report it as a candidate pending confirmation. If real, it would be the outermost and most temperate member.
3. The system is dynamically cold and near-resonant. SPOCK and $N$-body integrations indicate stability for $e \lesssim 0.05$, with eccentricity—not mass—being the limiting factor. Including the candidate further restricts the eccentricities.
4. The four objects, stepped in insolation (1.71–9.5 $S_\oplus$) around a single star, span the Earth–Venus transition range; two of them (c and d), validated, fall in the temperate regime under the equilibrium-temperature criterion, anchoring on confirmed planets the system's value as a comparative-planetology laboratory. However, their atmospheric retention is uncertain (it depends on the radius and the unconstrained past XUV flux) and the low TSM (~3–5) places the system beyond the reach of current atmospheric characterization.
5. TOI-789 exemplifies the numerically dominant population of compact terrestrial-planet systems around M dwarfs whose bright analog (L 98-59) has been characterized in detail, but which remain inaccessible to mass and atmosphere measurements. Their dynamical characterization offers a complementary route to constraining their properties.
6. Regardless of future prospects, the immediate value of the work is operational and methodological: the ephemerides refined over 42 sectors place the confirmation of the candidate .04 within reach of ground-based follow-up networks and the TESS extended mission; the analysis chain employed constitutes a reproducible template for compact systems around faint M dwarfs; and the system, together with the treatment of the ~28 days systematic and the in-pixel contamination dilution, provides reusable material for population studies and for the analysis of multisector light curves.

The confirmation of the fourth candidate and the refinement of the stellar parameters via dedicated spectroscopy are the natural steps to complete the characterization of the system.

## Acknowledgments

The author thanks Karen A. Collins and Richard P. Schwarz for their contributions to the TFOP-coordinated ground-based follow-up photometry and Karen A. Collins for helpful advice and recommendations. A large language model (Claude Opus 4.8, Anthropic) was used for translation, language improvement, and grammatical and syntactic editing of the manuscript text. All scientific content—the data reduction and analysis, the statistical validation, the dynamical modeling, and the interpretation and conclusions—is the author's own, and the author takes full responsibility for the accuracy and integrity of the work.

This paper includes data collected by the TESS mission, which are publicly available from the Mikulski Archive for Space Telescopes (MAST) at the Space Telescope Science Institute (STScI). Funding for the TESS mission is provided by NASA's Science Mission Directorate. We acknowledge the use of public TESS data from pipelines at the TESS Science Office and at the TESS Science Processing Operations Center. Resources supporting this work were provided by the NASA High-End Computing (HEC) Program through the NASA

Advanced Supercomputing (NAS) Division at Ames Research Center for the production of the SPOC data products. STScI is operated by the Association of Universities for Research in Astronomy, Inc., under NASA contract NAS 5–26555.

This research has made use of the Exoplanet Follow-up Observation Program (ExoFOP; DOI: 10.26134/ExoFOP5) website and the NASA Exoplanet Archive, which are operated by the California Institute of Technology, under contract with the National Aeronautics and Space Administration under the Exoplanet Exploration Program.

This work has made use of data from the European Space Agency (ESA) mission Gaia (https://www.cosmos.esa.int/gaia), processed by the Gaia Data Processing and Analysis Consortium (DPAC, https://www.cosmos.esa.int/web/gaia/dpac/consortium). Funding for the DPAC has been provided by national institutions, in particular the institutions participating in the Gaia Multilateral Agreement.

This publication makes use of data products from the Two Micron All Sky Survey, which is a joint project of the University of Massachusetts and the Infrared Processing and Analysis Center/California Institute of Technology, funded by the National Aeronautics and Space Administration and the National Science Foundation.

Based in part on observations obtained at the international Gemini Observatory, a program of NSF NOIRLab, which is managed by the Association of Universities for Research in Astronomy (AURA) under a cooperative agreement with the US National Science Foundation on behalf of the Gemini Observatory partnership. The Zorro speckle instrument was funded by the NASA Exoplanet Exploration Program and built at the NASA Ames Research Center by Steve B. Howell, Nic Scott, Elliott P. Horch, and Emmett Quigley.

Based in part on observations obtained at the Southern Astrophysical Research (SOAR) telescope, which is a joint project of the Ministério da Ciência, Tecnologia e Inovações (MCTI/LNA) do Brasil, the US National Science Foundation's NOIRLab, the University of North Carolina at Chapel Hill (UNC), and Michigan State University (MSU).

This work makes use of observations from the Las Cumbres Observatory global telescope network.

The MEarth Team gratefully acknowledges funding from the David and Lucile Packard Fellowship for Science and Engineering (awarded to D. Charbonneau). This material is based upon work supported by the National Science Foundation under grants AST-0807690, AST-1109468, AST-1616624, and AST-1616684, and the National Aeronautics and Space Administration under grant No. 80NSSC18K0476 issued through the XRP Program.

*Facilities:* TESS, Gemini:South (Zorro), SOAR (HRCam), LCOGT, MEarth, Gaia, 2MASS, Exoplanet Archive.

*Software:* `lightkurve` (Lightkurve Collaboration 2018), `transitleastsquares` (M. Hippke & R. Heller 2019), `TRICERATOPS` (S. Giacalone et al. 2021), `juliet` (N. Espinoza et al. 2019), `batman` (L. Kreidberg 2015), `dynesty` (J. S. Speagle 2020), `SPOCK` (D. Tamayo et al. 2020), `REBOUND` (H. Rein & S.-F. Liu 2012), `WHFast` (H. Rein & D. Tamayo 2015), `TTVFast` (K. M. Deck et al. 2014), `emcee` (D. Foreman-Mackey et al. 2013), Box Least Squares (G. Kovács et al. 2002), `numpy`, `scipy`, `matplotlib`, `astropy`.

## ORCID iDs

Juan F. Pagés Navarrete https://orcid.org/0009-0006-7910-4276

## References

Agol, E., Steffen, J., Sari, R., & Clarkson, W. 2005, MNRAS, 359, 567
Almenara, J. M., Bonfils, X., Bryant, E. M., et al. 2024, AJ, 167, 164
Andrae, R., Fouesneau, M., Sordo, R., et al. 2023, A&A, 674, A27
Angus, R., Morton, T., Aigrain, S., Foreman-Mackey, D., & Rajpaul, V. 2018, MNRAS, 474, 2094
Bailer-Jones, C. A. L., Rybizki, J., Fouesneau, M., Demleitner, M., & Andrae, R. 2021, AJ, 161, 147
Barclay, T., Pepper, J., & Quintana, E. V. 2018, ApJS, 239, 2
Batalha, N. M., Rowe, J. F., Gilliland, R. L., et al. 2010, ApJL, 713, L103
Brown, T. M., Baliber, N., Bianco, F. B., et al. 2013, PASP, 125, 1031
Burke, C. J., Bryson, S. T., Mullally, F., et al. 2014, ApJS, 210, 19
Chambers, J. E., Wetherill, G. W., & Boss, A. P. 1996, Icar, 119, 261
Chen, J., & Kipping, D. 2017, ApJ, 834, 17
Christiansen, J. L., Clarke, B. D., Burke, C. J., et al. 2015, ApJ, 810, 95
Christiansen, J. L., Clarke, B. D., Burke, C. J., et al. 2020, AJ, 155, 180
Cowan, N. B., Greene, T., Angerhausen, D., et al. 2015, PASP, 127, 311
Deck, K. M., Agol, E., Holman, M. J., & Nesvorny, D. 2014, ApJ, 787, 132
Delrez, L., Gillon, M., Queloz, D., et al. 2018, SPIE, 10700, 107001I
Dietrich, J., & Apai, D. 2020, AJ, 160, 107
Dietrich, J., Apai, D., & Malhotra, R. 2022, AJ, 163, 88
Dressing, C. D., & Charbonneau, D. 2015, ApJ, 807, 45
Espinoza, N., Kossakowski, D., & Brahm, R. 2019, MNRAS, 490, 2262
Fabrycky, D. C., Lissauer, J. J., Ragozzine, D., et al. 2014, ApJ, 790, 146
Fang, J., & Margot, J.-L. 2013, ApJ, 767, 115
Fetherolf, T., et al. 2023, MNRAS, 520, 5736
Foreman-Mackey, D., Hogg, D. W., Lang, D., & Goodman, J. 2013, PASP, 125, 306
Giacalone, S., & Dressing, C. D. 2020, Candidate exoplanet rating tool, Astrophysics Source Code Library, ascl:2002.004
Giacalone, S., Dressing, C. D., Jensen, E. L. N., et al. 2021, AJ, 161, 24
Gillis, E. D., Cloutier, R., & Pass, E. K. 2026, AJ, 171, 317
Gillon, M., Triaud, A. H. M. J., Demory, B.-O., et al. 2017, Natur, 542, 456
Gomez Barrientos, J., Greklek-McKeon, M., Knutson, H. A., et al. 2025, AJ, 170, 148
Greklek-McKeon, M., Gomez Barrientos, J., Knutson, H. A., et al. 2026, AJ, 171, 97
Guerrero, N. M., et al. 2021, ApJS, 254, 39
Günther, M. N., Pozuelos, F. J., Dittmann, J. A., et al. 2019, NatAs, 3, 1099
Heller, R., et al. 2019, A&A, 627, A66
Hippke, M., & Heller, R. 2019, A&A, 623, A39
Holman, M. J., & Murray, N. W. 2005, Sci, 307, 1288
Irwin, J. M., Berta-Thompson, Z. K., Charbonneau, D., et al. 2015, CSSS, 18, 767
Jenkins, J. M., Twicken, J. D., McCauliff, S., et al. 2016, SPIE, 9913, 99133E
Ji, X., Chatterjee, R. D., Coy, B. P., & Kite, E. S. 2025, ApJ, 992, 198
Kane, S. R., Kopparapu, R. K., & Domagal-Goldman, S. D. 2014, ApJL, 794, L5
Kawauchi, K., Murgas, F., Palle, E., et al. 2022, A&A, 666, A4
Kempton, E. M.-R., Bean, J. L., Louie, D. R., et al. 2018, PASP, 130, 114401
Kipping, D. M. 2013, MNRAS, 435, 2152
Kopparapu, R. K., Ramirez, R. M., SchottelKotte, J., et al. 2014, ApJL, 787, L29
Kostov, V. B., Schlieder, J. E., Barclay, T., et al. 2019, AJ, 158, 32
Kovács, G., Zucker, S., & Mazeh, T. 2002, A&A, 391, 369
Kreidberg, L. 2015, PASP, 127, 1161
Kunimoto, M., Lin, Z., Millholland, S., et al. 2025, AJ, 167, 174
Lafarga, M., Armstrong, D. J., Cui, K., et al. 2026, MNRAS, 548, stag512
Li, J., Tenenbaum, P., Twicken, J. D., et al. 2019, PASP, 131, 024506
Lightkurve Collaboration 2018, Lightkurve: Kepler and TESS time series analysis in Python, Astrophysics Source Code Library, ascl:1812.013
Lindegren, L., Klioner, S. A., Hernández, J., et al. 2021, A&A, 649, A2
Lissauer, J. J., Ragozzine, D., Fabrycky, D. C., et al. 2011, ApJS, 197, 8
Lissauer, J. J., Marcy, G. W., Rowe, J. F., et al. 2012, ApJ, 750, 112
Lithwick, Y., Xie, J., & Wu, Y. 2012, ApJ, 761, 122
López-Morales, M., Currie, T., Teske, J., et al. 2019, FrASS, 6, 42
Luque, R., Osborn, H. P., Leleu, A., et al. 2023, Natur, 623, 932
Mann, A. W., Brewer, J. M., Gaidos, E., Lépine, S., & Hilton, E. J. 2013, AJ, 145, 52

Mann, A. W., Feiden, G. A., Gaidos, E., Boyajian, T., & von Braun, K. 2015, ApJ, 804, 64
Mann, A. W., Dupuy, T., Kraus, A. L., et al. 2019, ApJ, 871, 63
McQuillan, A., Mazeh, T., & Aigrain, S. 2014, ApJS, 211, 24
Ment, K., & Charbonneau, D. 2023, AJ, 165, 265
Millholland, S., Wang, S., & Laughlin, G. 2017, ApJL, 849, L33
Mulders, G. D., Pascucci, I., Apai, D., & Ciesla, F. J. 2018, AJ, 156, 24
Nutzman, P., & Charbonneau, D. 2008, PASP, 120, 317
Obertas, A., Van Laerhoven, C., & Tamayo, D. 2017, Icar, 293, 52
Parashivamurthy, H. M., & Mulders, G. D. 2025, A&A, 703, A8
Pass, E. K., Charbonneau, D., & Vanderburg, A. 2025, ApJL, 986, L3
Pu, B., & Wu, Y. 2015, ApJ, 807, 45
Rein, H., & Liu, S.-F. 2012, A&A, 537, A128
Rein, H., & Tamayo, D. 2015, MNRAS, 452, 376
Ricker, G. R., Winn, J. N., Vanderspek, R., et al. 2015, JATIS, 1, 014003
Rodenbeck, K., Heller, R., Hippke, M., & Gizon, L. 2018, A&A, 617, A49
Rodler, F., & López-Morales, M. 2018, ApJ, 781, 54
Rojas-Ayala, B., Covey, K. R., Muirhead, B., & Lloyd, J. P. 2012, ApJ, 748, 93
Rowe, J. F., Bryson, S. T., Marcy, G. W., et al. 2014, ApJ, 784, 45
Scott, M. G., Dransfield, G., Timmermans, M., et al. 2026, MNRAS, 547, stag070
Scott, N. J., Howell, S. B., Gnilka, C. L., et al. 2021, FrASS, 8, 716560
Seager, S., & Mallén-Ornelas, G. 2003, ApJ, 585, 1038
Sebastian, D., Gillon, M., Ducrot, E., et al. 2021, A&A, 645, A100
Shields, A. L., Ballard, S., & Johnson, J. A. 2016, PhR, 663, 1
Smith, A. W., & Lissauer, J. J. 2009, Icar, 201, 381
Smith, J. C., Stumpe, M. C., Van Cleve, J. E., et al. 2012, PASP, 124, 1000
Snellen, I. A. G., de Kok, R. J., le Poole, R., Brogi, M., & Birkby, J. 2013, ApJ, 764, 182
Speagle, J. S. 2020, MNRAS, 493, 3132
Stassun, K. G., Oelkers, R. J., Paegert, M., et al. 2019, AJ, 158, 138
Stumpe, M. C., Smith, J. C., Catanzarite, J. H., et al. 2014, PASP, 126, 100
Sullivan, P. W., Winn, J. N., Berta-Thompson, Z. K., et al. 2015, ApJ, 809 77
Tamayo, D., Gilbertson, C., & Foreman-Mackey, D. 2021, MNRAS, 501, 4798
Tamayo, D., Rein, H., Petrovich, C., & Murray, N. 2017, ApJL, 840, L19
Tamayo, D., Cranmer, M., Hadden, S., et al. 2020, PNAS, 117, 18194
Thompson, S. E., Coughlin, J. L., Hoffman, K., et al. 2018, ApJS, 235, 38
Tokovinin, A. 2018, PASP, 130, 035002
Triaud, A. H. M. J., de Wit, J., Klein, F., et al. 2024, NatAs, 8, 17
Trotta, R. 2008, ConPh, 49, 71
Turtelboom, E. V., Dietrich, J., Dressing, C. D., & Harada, C. K. D. 2025, AJ, 170, 3
Twicken, J. D., Catanzarite, J. H., Clarke, B. D., et al. 2018, PASP, 130, 064502
Weiss, L. M., Marcy, G. W., Petigura, E. A., et al. 2018, AJ, 155, 48
Wood, M. L., et al. 2021, AJ, 162, 128
Wunderlich, F., Godolt, M., Grenfell, J. L., et al. 2019, A&A, 624, A49
Zahnle, K. J., & Catling, D. C. 2017, ApJ, 843, 122
Zechmeister, M., & Kürster, M. 2009, A&A, 496, 577